\documentclass{emulateapj}
\usepackage{apjfonts}
\usepackage{lscape}
\usepackage{amsmath}

\newcommand{\kt}{\ensuremath{k{\mathrm{T}}_{\mathrm{in}}}}

\newcommand{\Ledd}{\ensuremath{L_{\mathrm{Edd}}}}

\newcommand{\Lsol}{\ensuremath{L_{\mathrm{\odot}}}}

\newcommand{\LX}{\ensuremath{L_{\mathrm{X}}}}

\newcommand{\Msol}{\ensuremath{\mathrm{M}_{\odot}}}

\newcommand{\NH}{\ensuremath{N_{\mathrm{H}}}}

\newcommand{\arcs}{\ensuremath{^{\prime\prime}}}

\newcommand{\pcmsq}{\ensuremath{\cm^{-2}}}

\newcommand{\erg}{\ensuremath{\mbox{erg}}}

\newcommand{\cm}{\ensuremath{\mbox{cm}}}

\newcommand{\nm}{\ensuremath{\mbox{\nm}}}

\newcommand{\ps}{\ensuremath{\s^{-1}}}

\newcommand{\s}{\ensuremath{\mbox{s}}}

\newcommand{\ergps}{\ensuremath{\erg~\ps}}
\newcommand{\flux}{\ensuremath{\erg~\ps~\pcmsq}}

\newcommand{\etal}{{ et al.\thinspace}}

\newcommand{\CHANDRA}{\emph{Chandra}}

\newcommand{\XMM}{\emph{XMM-Newton}}

\newcommand{\chisq}{\ensuremath{\chi^2}}

\newcommand{\OIII}{[O~\textsc{iii}]}



\begin{document}

\title{The X-ray Spectra of the Luminous LMXBs in NGC 3379: Field and Globular Cluster Sources   \\}

\author{N. J. Brassington}
\affil{Harvard-Smithsonian Center for Astrophysics, 60 Garden Street, Cambridge, MA 02138}
\email{nbrassington@head.cfa.harvard.edu}

\author{G. Fabbiano}
\affil{Harvard-Smithsonian Center for Astrophysics, 60 Garden
Street, Cambridge, MA 02138}
\author{S. Blake}
\affil{School of Physics \& Astronomy, University of Southampton, Highfield, Southampton,
SO17 1BJ, U.K \\ \& Sub-Department of Astrophysics, University of Oxford, Oxford, OX1 3RH, UK}
\author{A. Zezas}
\affil{Harvard-Smithsonian Center for Astrophysics, 60 Garden
Street, Cambridge, MA 02138 \\ \& University of Crete, Physics Department, PO Box 2208, 710 03 Heraklion, Crete, Greece \\
   \& IESL, Foundation for Research and Technology, 71110 Heraklion, Crete, Greece }

\author{L. Angelini}
\affil{Laboratory for X-Ray Astrophysics, NASA Goddard Space Flight Center, Greenbelt, MD 20771}
\author{R. L. Davies}
\affil{Sub-Department of Astrophysics, University of Oxford, Oxford, OX1 3RH, UK}
\author{J. Gallagher}
\affil{Department of Astronomy, University of Wisconsin, Madison, WI 53706-1582}
\author{V. Kalogera}
\affil{Department of Physics and Astronomy, Northwestern University, Evanston, IL 60208}
\author{D.-W. Kim}
\affil{Harvard-Smithsonian Center for Astrophysics, 60 Garden
Street, Cambridge, MA 02138}
\author{A. R. King}
\affil{Theoretical Astrophysics Group, University of Leicester, Leicester 
LE1 7RH, UK}
\author{A. Kundu}
\affil{Department of Physics and Astronomy, Michigan State University, East Lansing, MI 48824-2320}
\author{G. Trinchieri}
\affil{INAF-Osservatorio Astronomico di Brera, Via Brera 28, 20121 Milan, Italy}
\author{S. Zepf}
\affil{Department of Physics and Astronomy, Michigan State University, East Lansing, MI 48824-2320}

\shorttitle{Luminous LMXBs in NGC 3379}
\shortauthors{Brassington \etal}
\bigskip

\begin{abstract}

From a deep multi-epoch \CHANDRA\ observation of the elliptical galaxy NGC 3379 we report the spectral properties of eight luminous LMXBs (\LX$>1.2\times10^{38}$\ergps). We also present a set of spectral simulations, produced to aid the interpretation of low-count single-component spectral modeling. These simulations demonstrate that it is possible to infer the spectral states of X-ray binaries from these simple models and thereby constrain the properties of the source.

Of the eight LMXBs studied, three reside within globular clusters, and one is a confirmed field source. Due to the nature of the luminosity cut all sources are either neutron star binaries emitting at or above the Eddington luminosity or black hole binaries. The spectra from these sources are well described by single-component models, with parameters consistent with Galactic LMXB observations, where hard-state sources have a range in photon index of 1.5$-$1.9 and thermally dominant sources have inner disc temperatures between $\sim0.7-1.55$ keV. 

The large variability observed in the brightest globular cluster source (\LX$>4\times10^{38}$\ergps) suggests the presence of a black hole binary. At its most luminous this source is observed in a thermally dominant state with \kt=1.5 keV, consistent with a black hole mass of $\sim$4\Msol. This observation provides further evidence that globular clusters are able to retain such massive binaries.
We also observed a source transitioning from a bright state (\LX$\sim 1\times10^{39}$\ergps), with prominent thermal and non-thermal components, to a less luminous hard state (\LX=3.8$\times10^{38}$\ergps, $\Gamma=$1.85). In its high flux emission this source exhibits a cool-disc component of $\sim$0.14 keV, similar to spectra observed in some ultraluminous X-ray sources. Such a similarity indicates a possible link between `normal' stellar mass black holes in a high accretion state and ULXs.
%


\end{abstract}

\section{Introduction}

The discovery with \CHANDRA\ of several low-mass X-ray binary (LMXB) populations in early-type galaxies, and the associations of these LMXBs with either Globular Clusters (GCs) or the stellar field, have provided new impetus to the study of the formation and evolution of  LMXBs in GCs and to the possible relation of field LMXBs to the GC population (Grindlay \& Hertz 1985; Verbunt \& van den Heuvel 1995; see Fabbiano 2006 and refs. therein). Given the characteristics of these data, most of this work has been based on population studies (e.g., Kim \etal\ 2006: Kundu \etal\ 2007; Sivakoff et al 2007; Voss \& Gilfanov 2007). However, in the few cases of detection of luminous sources in deep enough observations, detailed spectral and variability studies can be pursued, to provide more direct constraints on the nature of the X-ray sources. One such example is the recent discovery of a variable luminous GC source in NGC 4472, with temporal and spectral characteristics supporting a stellar BH binary (Maccarone et al 2007; Shih et al 2008).

In this paper we report the results of the spectral analysis of eight luminous sources (\LX $> 1.2\times 10^{38}$ \ergps\ in the 0.3$-$8.0 keV band), detected in the nearby elliptical galaxy NGC 3379 (in the poor group Leo, D=10.6 Mpc, Tonry \etal\ 2001\footnote{Note that Jensen \etal\ (2003) report a distance of 9.82 Mpc. However, the distance of Tonry \etal\ (2001) is adopted here for consistency with Brassington \etal\ (2008). This choice does not affect our conclusions in any way.}) with \CHANDRA\ ACIS-S (Weisskopf et al 2000). These sources were observed at five different epochs, as part of a monitoring campaign with \CHANDRA\ (PI: Fabbiano) providing the rare opportunity of long-term spectral monitoring of LMXBs in an elliptical galaxy. These sources are part of the sample of 132 sources detected in NGC 3379 from these observations with luminosities greater than a few 10$^{36}$ \ergps\ (Brassington \etal\ 2008 $-$ hereafter B08).

In section \ref{sec:selection} we describe the observations and the properties of the sources under study from the B08 catalog. In section \ref{sec:results} we describe our spectral analysis and report the results. In section \ref{sec:sims} we present spectral simulations that we have performed to aid in the interpretation of our results and in section \ref{sec:discussion} we discuss the spectral analysis and compare the results to our simulations. Our conclusions are summarized in section \ref{sec:conclusions}.

\section{Source Selection}
\label{sec:selection}

From the catalog of B08 we have selected a sub-set of the brightest sources (\LX $>1.2\times 10^{38}$ \ergps) within the D$_{25}$ ellipse of the galaxy to perform detailed spectral and temporal analysis. This luminosity cut results in the selection of 12 sources, we further exclude the nuclear source (S81) and the confirmed ULX (S70; Fabbiano \etal\ 2006), both of which will be the subject of forthcoming papers. We also exclude confused sources in the central region (S75 \& S82), where reliable spectral extraction cannot be carried out. This results in a final selection of eight sources. 
Of these, six lie within the {\em HST} WFPC2 field of view of NGC 3379, allowing firm constraints to be placed on their optical counterparts (see B08). Three of these sources (S41, S42 \& S67 of B08) are identified with GCs and one has been confirmed as a field LMXB (S86). The other two sources are found within the region of confusion from the {\em HST} observation (S74 \& S77) and are therefore unclassified. The two remaining sources are external to the optical FOV (S102 \& S103) and are therefore also unclassified. At a flux of 8.9$\times10^{-15}$\flux\ (which gives \LX=1.2$\times10^{38}$ \ergps\ at D=10.6 Mpc), from the $ChaMP+CDF\ logN-logS$ relation of Kim \etal\ (2007), less than 1 source is expected to be a background AGN over the area encompassing the eight sources. The \LX$>1.2\times10^{38}$ \ergps\ luminosity cut ensures that the selected sources have a minimum of $\sim$400 counts from the co-added observation and 100 counts in at least one individual observation, allowing meaningful spectral analysis to be carried out. By nature of the selection criteria these sources are either NS binaries emitting near or above the Eddington luminosity, or are black hole binaries. 

From the observations of NGC 3379 the parameters derived in B08 indicate that BH-LMXBs are present in the X-ray source population of this galaxy, both in the field and in GCs. BH-LMXBs in the field are a well-established class of X-ray binaries in the Galaxy (see the review of Remillard \& McClintock 2006; from hereon in RM06). Although BH-LMXBs are not expected to be common or luminous in GCs (Kalogera \etal\ 2004; see Ivanova \etal\ 2010 for a recent discussion of the formation of BHs within GCs), there is recent convincing evidence of at least two such luminous sources. The first source, a ULX detected within the elliptical Virgo galaxy NGC 4472, exhibits strong X-ray variability (Maccarone \etal\ 2007, Shih \etal\ 2008) providing unambiguous evidence for a BH-GC. Subsequent optical spectroscopy of the GC has revealed very broad \OIII\ emission lines, leading to the suggestion that this source is a stellar mass black hole accreting above or near the Eddington limit (Zepf \etal\ 2008). The second source, a ULX in NGC 1399 (Irwin \etal\ 2010), also exhibits strong \OIII\ emission lines (although these are less broad than the NGC 4472 source) and has little or no hydrogen emission. This source could also be a stellar mass black hole accreting at or near the Eddington limit, although Irwin \etal\ (2010) suggest that tidal disruption of a white dwarf from an intermediate-mass black hole (IMBH: 100$-10^4$ \Msol) could explain the optical emission of this GC. From this interpretation a minimum black hole mass of 1000 \Msol\ is implied. 
If sources S41, S42 \& S67 are luminous BH binaries, the estimated a-posterior probability of a BH-GC association is $\sim$4\%, given that 70 GCs are found in NGC 3379 within the joint Chandra/Hubble field of view (see B08 and refs. therein). Given that only 10 GCs are associated with X-ray sources in NGC 3379 (for \LX\ $>$ a few 10$^{36}$ \ergps; B08 and Kim \etal\ 2009), these luminous BH-LMXBs may constitute a significant fraction of the GC-LMXB population. 

\begin{figure}
\begin{centering}
  \includegraphics[height=\linewidth,angle=-90]{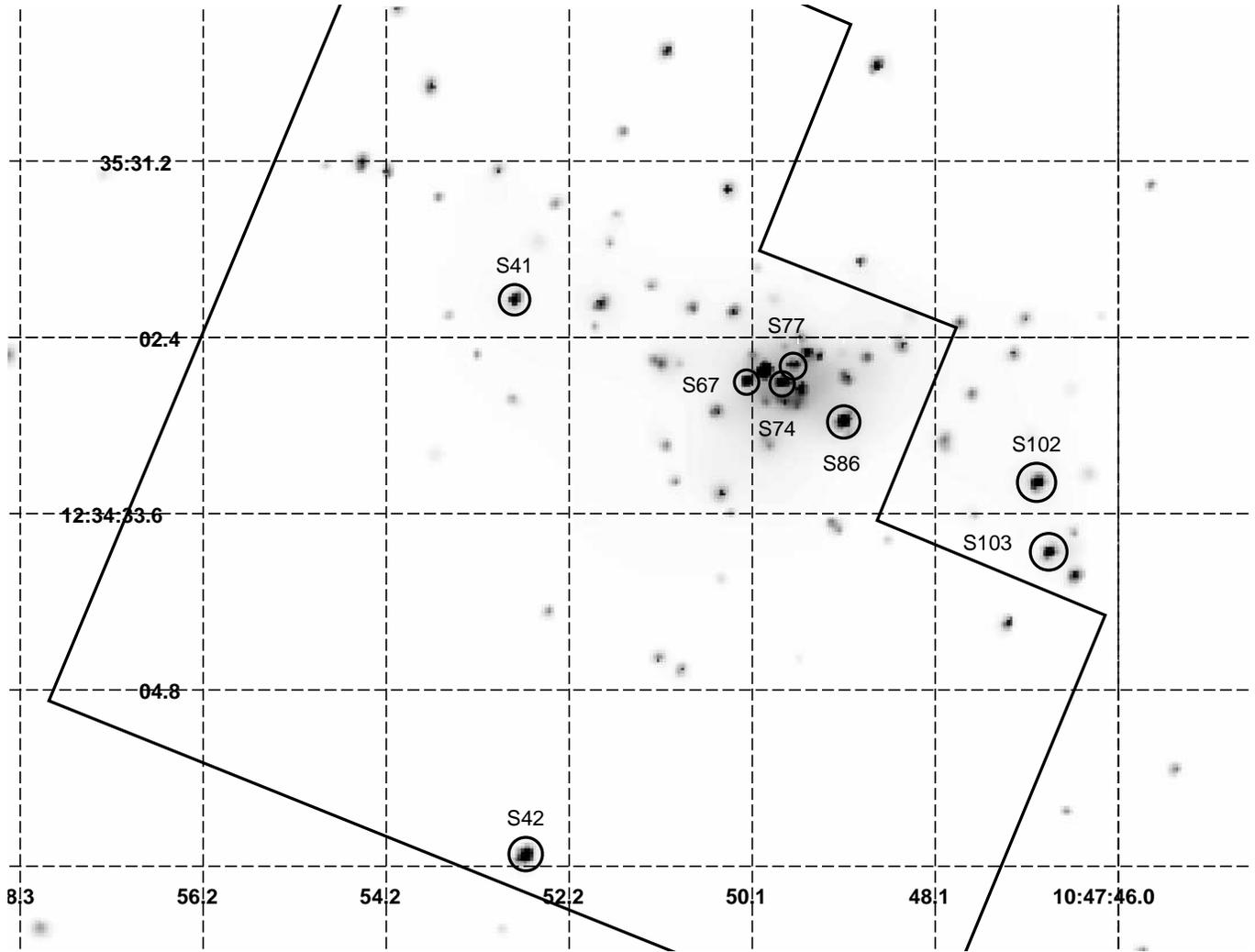}
\caption{Adaptively smoothed full-band (0.3$-$8.0 keV) X-ray image of NGC 3379. Overlaid in this figure are labeled circular regions indicating the eight source presented in this paper and {\em HST} FOV.}\label{fig:reg}
\end{centering}
\end{figure}

The eight sources presented here are shown in Figure \ref{fig:reg}, where an adaptively smoothed 0.3$-$8.0 keV X-ray image is presented, with the source regions and the {\em HST} FOV overlaid.
In Table \ref{tab:log} the log of the \CHANDRA\ observations is reported, along with the background-subtracted net source counts in the 0.3$-$8.0 keV band, for each source in each observation (as reported in B08). In this table the optical correlations are also identified, where GC indicates a confirmed globular cluster counterpart, F a field sources and - indicates sources with insufficient optical data to confirm a counterpart (or lack thereof). 

Figure \ref{fig:lc} (adapted from B08) shows the light curves of the eight sources from the \CHANDRA\ observations with panels indicating X-ray luminosity, the hardness ratio (HR) and color values of the source in each observation. In the top panel \LX\ indicates the X-ray luminosity in the 0.3$-$8.0 keV band, for an assumed power law spectrum with $\Gamma$=1.7 and Galactic line-of-sight\footnote{\NH=2.78$\times10^{20}$cm$^{-2}$ calculated with the tool COLDEN: \\ http://cxc.harvard.edu/toolkit/colden.jsp}, where the horizontal dashed line indicates the \LX\ derived from the co-added observation. Because the same canonical model has been applied to each source this value represents a scaling of the extracted source count-rate, where both the spatial and the temporal quantum efficiency variations have been accounted for by generating an energy conversion factor for each sources in each observation (see \textsection 2.1 in B08 for more details). 
In the second panel, the HR variation is shown where HR=(Hc-Sc)/(Hc+Sc), with Sc and Hc net counts in the 0.5-2.0 keV and 2.0-8.0 keV bands respectively. In the third and fourth panels the X-ray colors C21 and C32 are presented, these colors are defined as C21~=~log(S$_1$/S$_2$) and C32~=~log(S$_2$/H), where S$_1$, S$_2$ and H are the net counts respectively in the energy bands of 0.3$-$0.9 keV, 0.9$-$2.5 keV and 2.5$-$8.0 keV. 
As can be seen from these figures, the eight sources are persistent, in that they were detected with comparable luminosity over a five year span, although the majority exhibits some variability in luminosity and/or spectral properties. 

In Table \ref{tab:var} the long- and short-term variability of the sources are summarized, where the short-term variability is defined for each pointing, by means of the Kolmogorov-Smirnov test (K-S test). Here V indicates variable sources (with values $>$99\% confidence), P indicates possible variable sources (with variability values $>$90\% confidence) and the - indicates non-variable sources. The long-term variability from all five observations was defined by the chi-squared test where V indicates variability, and N indicates a non-variable source. In the last column, {\em Sign} presents the significance in change in luminosity between the highest and lowest flux observations. All of these classification are fully described in B08, \textsection 2.4. All the sources, with the exception of S67, have been determined to exhibit long-term variability. 

\section{Spectral Analysis and Results}
\label{sec:results}

The data were processed and calibrated as described in B08. Analysis was performed with the CXC CIAO software suite (v3.4)\footnote{http://asc.harvard.edu/ciao} with CALDB version 3.5 and HEASOFT (v5.3.1). Spectra were extracted for every pointing for each of the eight sources, using the CIAO tool {\em psextract}, where source regions were defined to be circular regions with radii as in B08; background counts were extracted from surrounding annuli, with outer radii 2-3 times larger, depending on the presence of nearby sources. In cases where the source region was found to overlap with a nearby sources, the extractions region radius was reduced and the area of the overlapping source was excluded. The minimum defined extraction radius was 1.5\arcsec\ on-axis, and 3\arcsec\ off-axis (maximum off-axis angle $\sim$1.5\arcmin), ensuring that the source regions enclosed $>$93\% of the encircled energy in all cases.

The source spectra were fitted in XSPEC (v11.3), where the data were restricted to 0.3$-$8.0 keV, as energies below this have calibration uncertainties, and the spectra presented here do not have significant source flux above 8.0 keV, where the data are highly contaminated by cosmic rays. The spectra were fitted to two models that describe the properties of X-ray binary spectra well (see e.g. review by RM06): the multicolor disc blackbody (DISKBB in XSPEC; hereon in referred to as DBB), and the power law (PO) models, along with the $wabs$ photoelectric absorption model\footnote{Given the statistics of our data and the ACIS resolution, use of other absorption models does not affect our results and conclusions (see \textsection\ref{subsec:absorp}).}. All parameters were allowed to vary freely, with the exception of the absorption column \NH, which in cases where the best-fit value was below that of the Galactic absorption, was frozen to the Galactic value of 2.78$\times 10^{20}$\pcmsq.
In instances where there were sufficient counts to bin the data to at least 20 counts per bin (allowing Gaussian error approximation to be used), the minimum \chisq\ method was used to fit the data. Where there were too few counts, the Cash statistic (Cash 1979) was used in preference to \chisq, where the spectra were extracted without binning. However, this statistical method has the disadvantage that it does not provide a goodness-of-fit measure like \chisq.

Spectra for each source were first fitted separately for each pointing. These individual best-fit models, alongside the \LX\ and HR behavior shown in Figure \ref{fig:lc}, were then used to guide the joint fits for each source. In Table \ref{tab:Bestfit} these co-added fits are summarized, where column (1) gives the source number, column (2) the observations used in each fit, column (3) the net source counts, column (4) indicates which spectral model was used, column (5) the fit statistic (\chisq\ and number of degrees of freedom $\nu$, or the C statistics - indicated by $C$), column (6) the null hypothesis probability (or goodness when using the Cash statistic). Columns (7), (8) \& (9) present the best-fit values of the fit parameters with 1$\sigma$ errors for 1 interesting parameter: \NH\ (2.8Fr denotes that the value was frozen at Galactic column density), $\Gamma$ for the PO model and \kt, the temperature of the innermost stable orbit of an accretion disc, for the DBB. Column (10) indicates the intrinsic value of \LX\ (calculated as a weighted average of the individual luminosities from the joint fit) and column (11) the source luminosity range. This range was taken from the (1$\sigma$) lower-bound value of the lowest flux observation included in the joint spectral fit, to the (1$\sigma$) upper-bound value of the highest flux observation included in the joint fit. When only one spectrum was modeled, the range indicates the lower- and upper-bound \LX\ values from that single observation.

\subsection{Globular Cluster Sources}
\label{sec:GC}

Of the sub-sample of sources presented in this paper, three have been determined to be GC-LMXBs. Through inspection of their \LX, HR and color values presented in Figure \ref{fig:lc}, and the individual spectral fits of each source, two LMXBs (S41 \& S67) have been determined to exhibit little flux and no spectral variability, whilst the remaining source (S42) has been found to exhibit spectral changes between each observation, with a variance in flux significance of 6$\sigma$. In the case of the non-varying LMXBs, the five spectra for each source were jointly fitted in XSPEC, where all parameters, apart from the normalizations, were linked. For both sources the DBB and PO models were fitted to the spectra, these best-fits are summarized in Table \ref{tab:Bestfit}. In cases where the DBB model was statistically rejected the PO model only is presented. 

\begin{figure}
  \centering
	\begin{minipage}{0.45\linewidth}
	\centering

    \includegraphics[height=\linewidth,angle=-90]{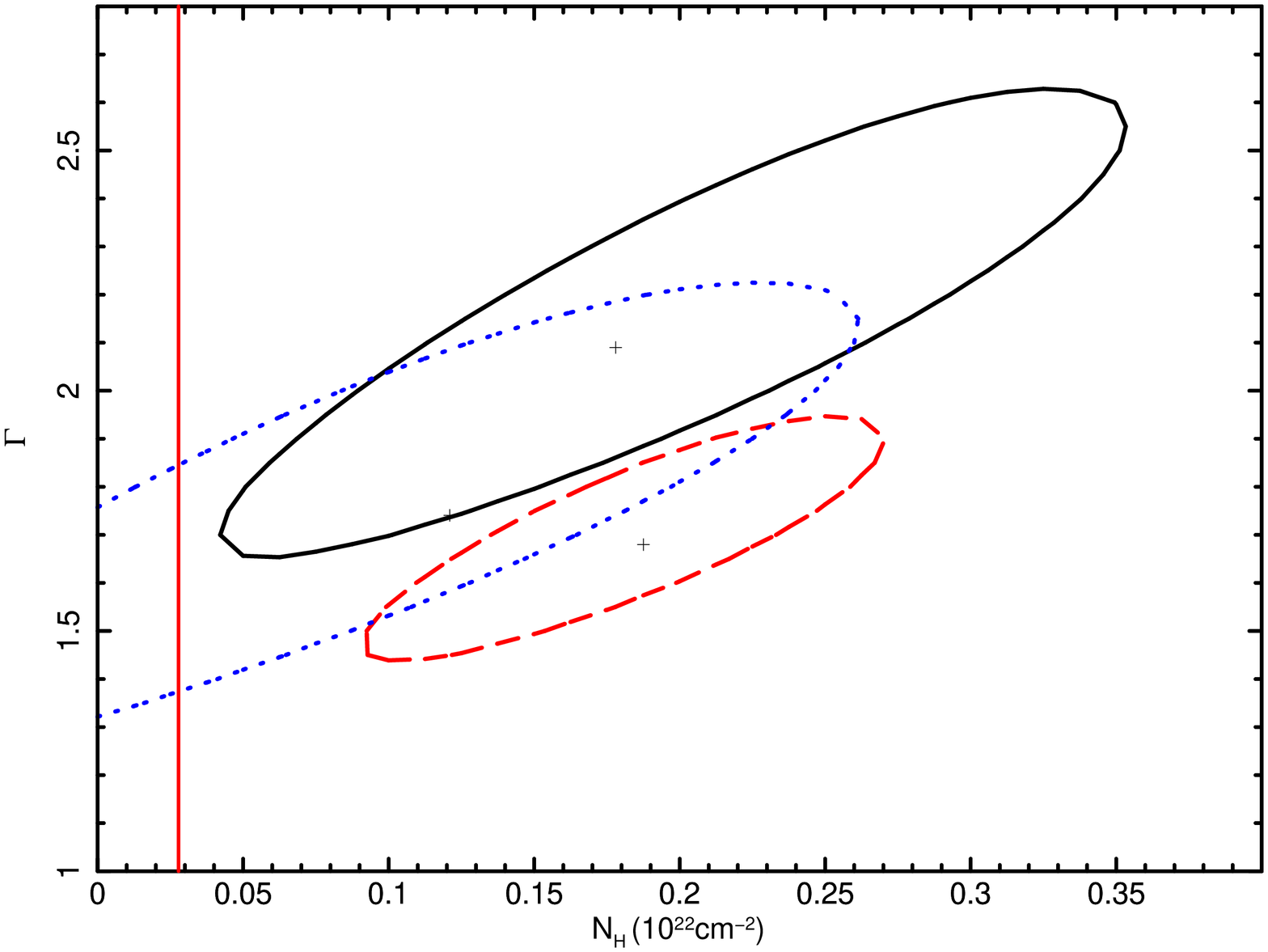}

	\end{minipage}
	\begin{minipage}{0.45\linewidth}
	\centering

    \includegraphics[height=\linewidth,angle=-90]{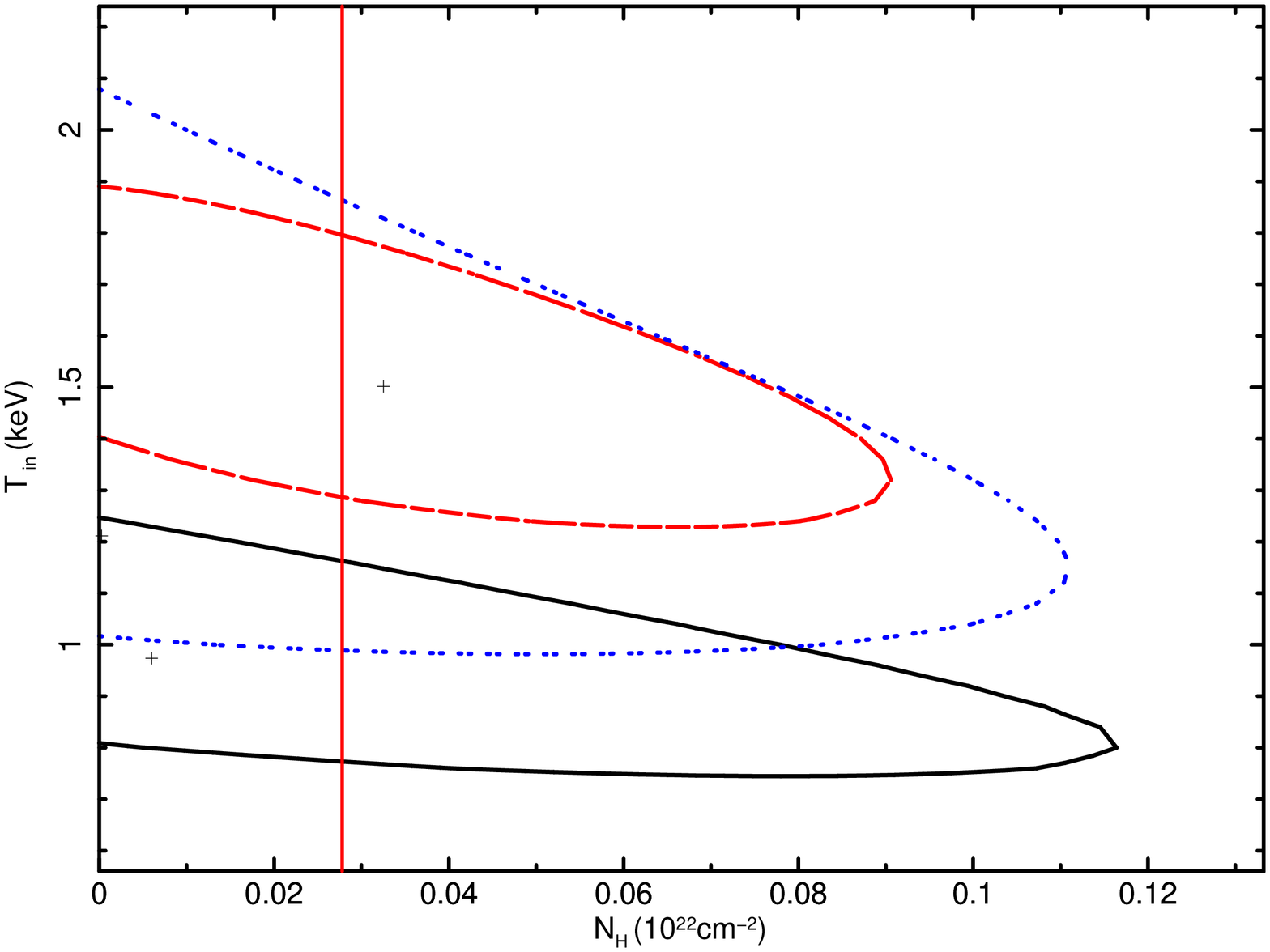}

	\end{minipage}
	\centering

    \caption{Contours for PO (left) and DBB fits (right) for S42. In both figures the black solid line represents the 2$\sigma$ contour for observations 1 \& 3, the red dashed line observations 2 \& 4 and the blue dotted line observation 5. Best-fit values are indicated by the cross hair within each contour (these values differ from the best-fit parameters presented in Table \ref{tab:Bestfit} when \NH\ is below Galactic absorption). In some instances cross hairs are not readily observable, in these cases the best-fit value of \NH\ tended to 0 when left free to vary. The solid vertical line indicates Galactic \NH. From the simulations presented in section \ref{sec:sims} the spectral properties of S42 suggest that this source is predominately in a thermally dominant state, entering an intermediate state in observation 5 (see text for details). \label{fig:Src42con} }

\end{figure}

For the variable source, S42, the most luminous GC-LMXB in our sample, the individual fits and \LX\ and HR values from B08 indicate that the source has been observed in three different spectral states. The spectra have therefore been grouped to reflect these differences. The first grouping contains spectra with lower \LX\ and soft HR value (obs 1 \&3), the second, spectra with higher \LX\ and hard HR value (obs 2\&4) and the remaining spectrum indicates an intermediate state (obs 5). Both the DBB and PO models were applied to these three joint fits and are summarized in Table \ref{tab:Bestfit}. The 90\% confidence contours for two interesting parameters are presented in Figure \ref{fig:Src42con} where the left-hand panel shows the PO model in the three spectral groups and right-hand panel the DBB fits. In both panels Galactic \NH\ is indicated by the vertical solid line.

\subsection{The Field Source}

\begin{figure}
\begin{centering}
  \includegraphics[width=0.5\linewidth]{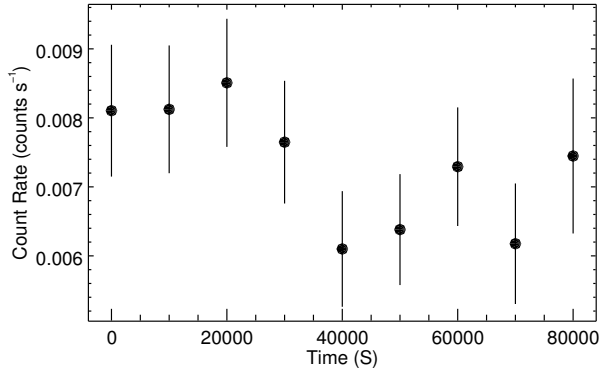}
\caption{Short-term lightcurve of S86 from observation 2. The x-axis provides time in seconds, beginning from the start of the observation, and the y-axis provides count-rate in count s$^{-1}$ between 0.3$-$8.0 keV. Binning of 10000 sec was used when extracting counts to provide adequately constrained count-rates to identify variability. }\label{fig:lc86}
\end{centering}
\end{figure}

Only one source presented in this paper has been confirmed as a field-LMXB, S86. It has been determined that this source not only exhibits long-term variable behavior in both its flux and spectra, but that it also displays short-term flux variability (defined as a possible variable source with the K-S test: Table \ref{tab:var}) during the second pointing. This short-term variability has been further investigated through examining the light-curve of the source during observation 2, which was created with the CIAO tool {\em dmextract}. To provide well constrained values of count-rate, allowing variability to be identified, binning of 10000 s was used when extracting counts. This resulting light-curve is shown in Figure \ref{fig:lc86}, where it can be seen that the count-rate drops from $\sim$0.008 cnt s$^{-1}$ to $\sim$0.006 cnt s$^{-1}$ part-way through the observation. Spectra were extracted from these two different count-rate epochs, to identify if a change in spectral shape accompanied this change in flux, however, it was determined that, within errors, the best-fit values from the two spectra were identical.

\begin{figure}
  \centering
	\begin{minipage}{0.45\linewidth}
	\centering

    \includegraphics[height=\linewidth,angle=-90]{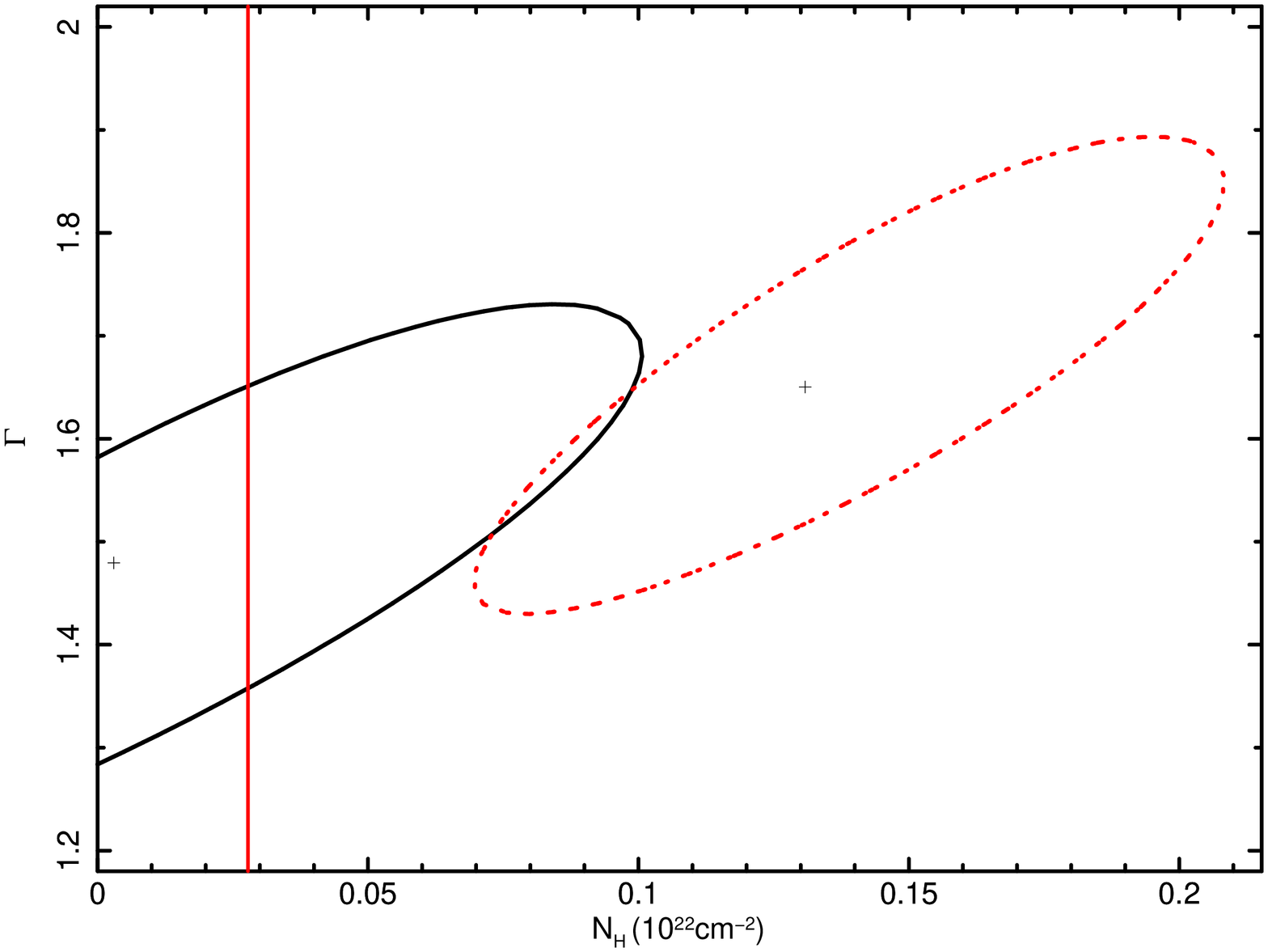}

	\end{minipage}
	\begin{minipage}{0.45\linewidth}
	\centering

    \includegraphics[height=\linewidth,angle=-90]{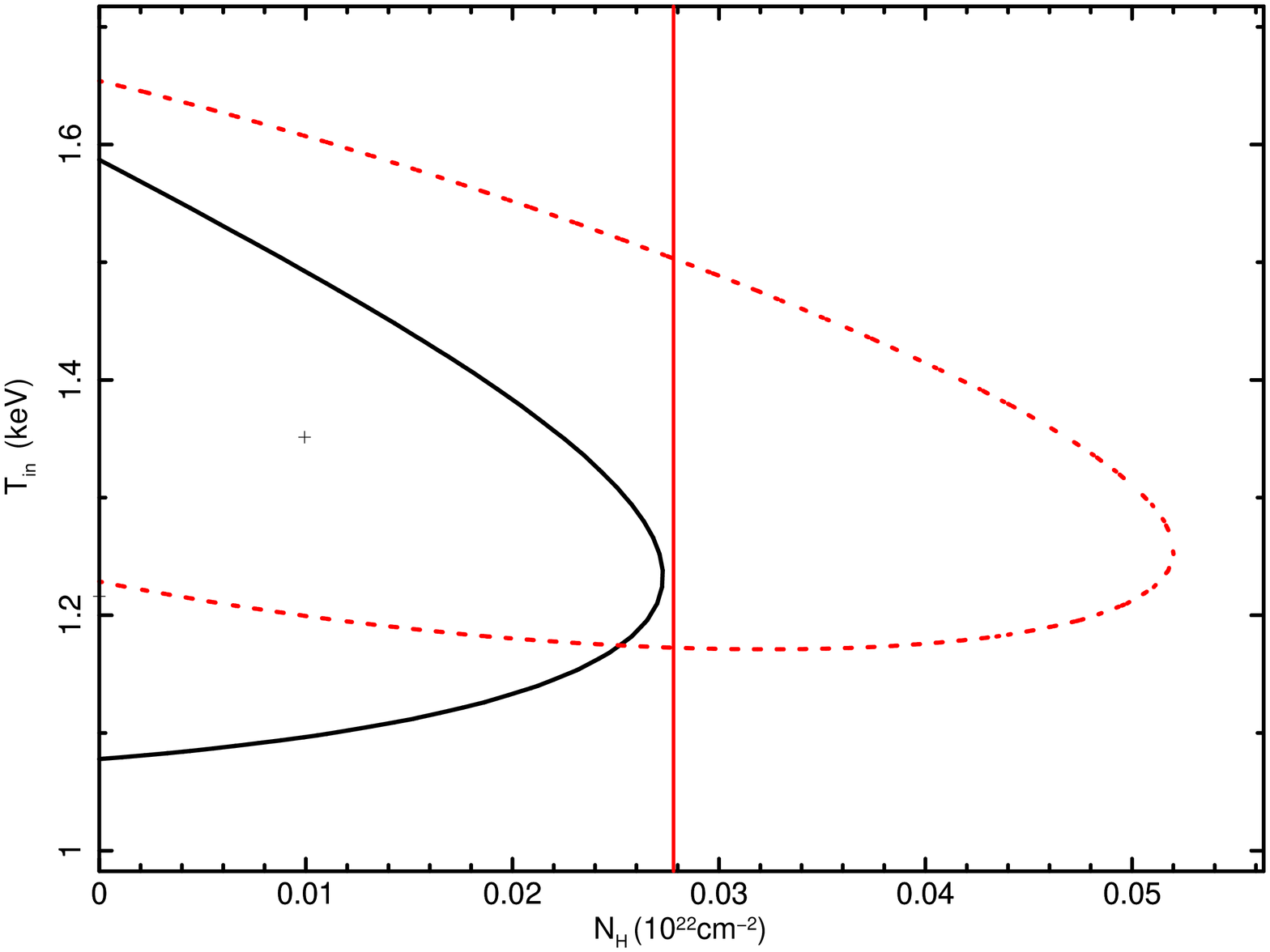}

	\end{minipage}
	\centering

    \caption{Contours for PO (left) and DBB fits (right) for S86. In both figures the black solid line represents the 2$\sigma$ contour for observations 1,3 \& 5, the red dotted line observations 2 \& 4. Best-fit values are indicated by the cross hair within each contour (these values differ from the best-fit parameters presented in Table \ref{tab:Bestfit} when \NH\ is below Galactic absorption). In some instances cross hairs are not readily observable, in these cases the best-fit value of \NH\ tended to 0 when left free to vary. The solid vertical line indicates Galactic \NH. From the simulations presented in section \ref{sec:sims} the spectral properties of S86 indicate that this source is transitioning between a hard state (or possibly a cool-disc state) to a thermally dominant state (see text for details). \label{fig:Src86con} }

\end{figure}

Once it had been determined that the short-term variability of S86 did not correspond to any spectral variation, the long-term spectral variability was investigated. Once again, guided by the individual spectral fits and the X-ray luminosity and HR values, the spectra were separated into two groups; a low \LX\ hard HR state (1,3\&5) and a high \LX\ softer HR state (2\&4). These best-fit values are summarized in Table \ref{tab:Bestfit} and their confidence contours are presented in Figure \ref{fig:Src86con}.

\subsection{Unclassified Sources}

The remaining four sources in this paper have insufficient optical data to confirm them as field or GC sources, either residing with the central 5\arcs\ of the galaxy, where source confusion means GC sources cannot be easily identified, or are detected outside of the {\em HST} FOV. Of these sources, two (S77 \& S103) have been determined to have no spectral variability, despite both of them being being classified as having flux variability. Consequently all five observations were fitted jointly for each source, following the methods described in \textsection \ref{sec:GC} for S41 \& S67.

Of the two remaining LMXBs (S74 \& S102), both have been determined to have variability and require separate spectral modeling. S74 exhibits stable long-term flux and spectra over the first four observations but in observation five a four-fold increase in flux (with a significance in flux change of 16$\sigma$) and a softening in HR is observed. The joint fits have therefore been separated into obs 1,2,3\&4, and observation 5, to reflect these variations. 
When modeling the spectrum from observation 5 it was found that there was a drop in count-rate in the binned data at $\sim$1.15 keV and from additional analysis it was determined that this feature was still present in the unbinned data. To further investigate this, additional components were included in the spectral model in an attempt to describe this feature. However, no physical models described the data well. Background subtraction, edge effects and calibration uncertainties were also considered as an explanation of the low energy bin but after further investigation all were rejected. After ruling these possibilities out, the most likely remaining explanation of this `feature' is that it is a consequence of statistics, and not a physical feature of the source. The PO and DBB models were fitted to the whole of the 0.3$-$8.0 keV spectrum and these best-fit model parameters are presented in Table \ref{tab:Bestfit}. In subsequent fits the energy range of 1.1$-$1.2 keV was excluded to determine the effect of the low-count bin. From this it was shown that the best-fit values are consistent with those from the full spectrum fit, although the goodness-of-fit values are greatly improved by excluding this energy range (PO \chisq/$\nu$ from 1.69 to 1.30; DBB \chisq/$\nu$ from 1.62 to 1.26).
Some short-term (possible) variability was also observed in observation 1 of S74, although there are too few counts in this pointing to be able to further investigate this. 

Through inspection of not only the long-term light-curve and HR values of S102 (Figure \ref{fig:lc}) but also its individual fits, there is a suggestion of a gradual and continual change in flux. Initially this is an increase in flux from the archival observation to the more recent pointings, followed by a steady decline, which is coupled with a suggestion of hardening in HR values. Therefore the joint fits were divided into three groups. Observation 1, the softest HR value, observations 2\&3, high \LX, and observations 4\&5, lower \LX. As with the previous sources, single-component PO and DBB models were applied to the spectra. 

However, for group 2, when the source count-rate is at its highest, the best-fit values from the PO model did not provide physically realistic parameters where the best-fit value of \NH\ was significantly below Galactic \NH, therefore a composite DBB+PO model was applied to the data.

\begin{figure}
  \centering
	\begin{minipage}{0.45\linewidth}
	\centering

    \includegraphics[height=\linewidth,angle=-90]{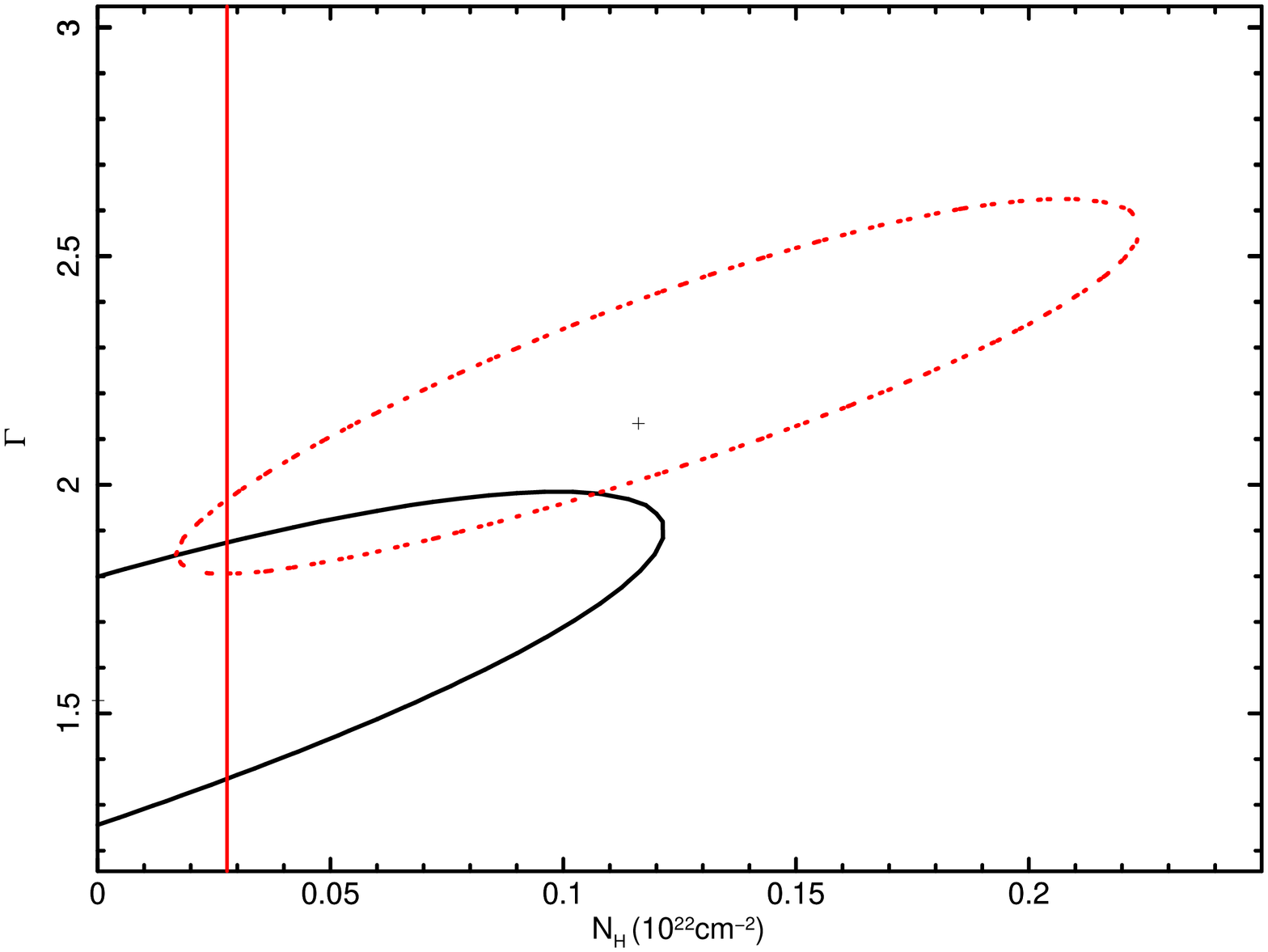}

	\end{minipage}
	\begin{minipage}{0.45\linewidth}
	\centering

    \includegraphics[height=\linewidth,angle=-90]{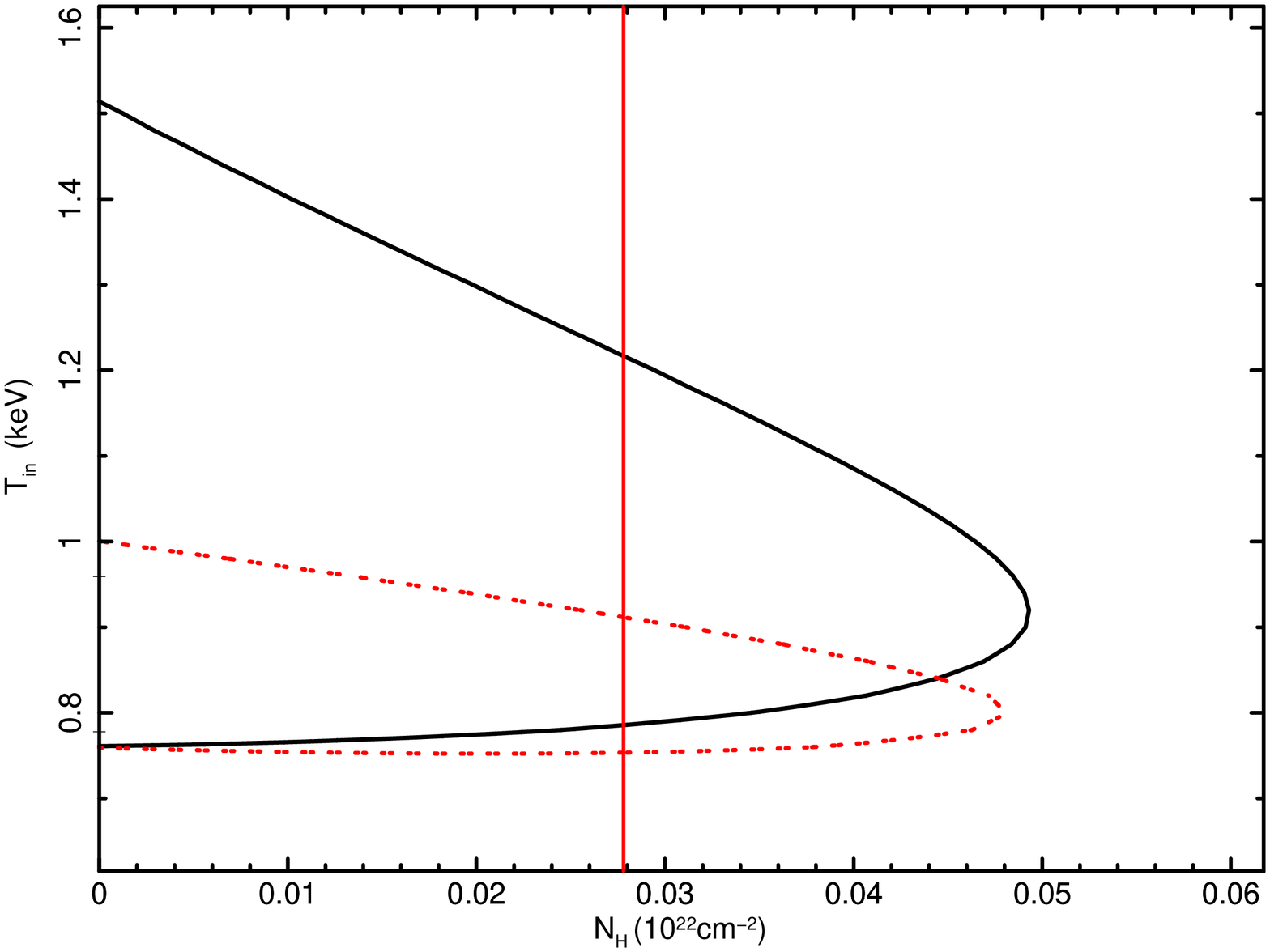}

	\end{minipage}
	\centering

    \caption{Contours for PO (left) and DBB fits (right) for S74. In both figures the black solid line represents the 2$\sigma$ contour for observations 1,2,3 \& 4, the red dotted line observations 5. Best-fit values are indicated by the cross hair within each contour (these values differ from the best-fit parameters presented in Table \ref{tab:Bestfit} when \NH\ is below Galactic absorption). In some instances cross hairs are not readily observable, in these cases the best-fit value of \NH\ tended to 0 when left free to vary. The solid vertical line indicates Galactic \NH. Based on the simulations presented in section \ref{sec:sims}, S74 is likely to be in a hard state in observations 1$-$4, transitioning to a thermally dominant state in observation 5 (see text for details). \label{fig:Src74con} }

\end{figure}

\begin{figure}
  \centering
    \includegraphics[angle=-90,width=0.45\linewidth]{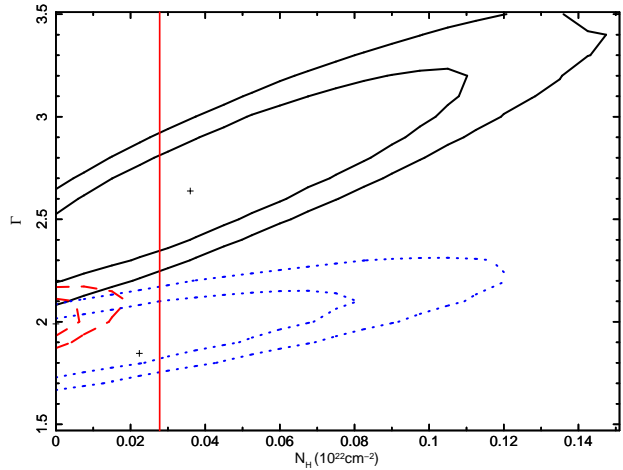}
    \caption{1$\sigma$ and 2$\sigma$ contours for the PO best-fit for S102. The solid black lines represents contours for observation 1, the red dashed lines show contours for observations 2 \& 3 and the dotted blue lines observations 4 \& 5. Best-fit values are indicated by the cross hair within each contour (these values differ from the best-fit parameters presented in Table \ref{tab:Bestfit} when \NH\ is below Galactic absorption). In some instances cross hairs are not readily observable, in these cases the best-fit value of \NH\ tended to 0 when left free to vary. The solid vertical line indicates Galactic \NH. From the simulations presented in section \ref{sec:sims}, S102 is likely to initially be in a steep power law state, transitioning to a luminous cool-disc state in observations 2 \& 3 and then changing to a hard state in observations 4 \& 5 (see text for details). \label{fig:Src102con} }

\end{figure}

\begin{figure}
  \centering
	\begin{minipage}{0.45\linewidth}
	\centering

    \includegraphics[height=\linewidth,angle=-90]{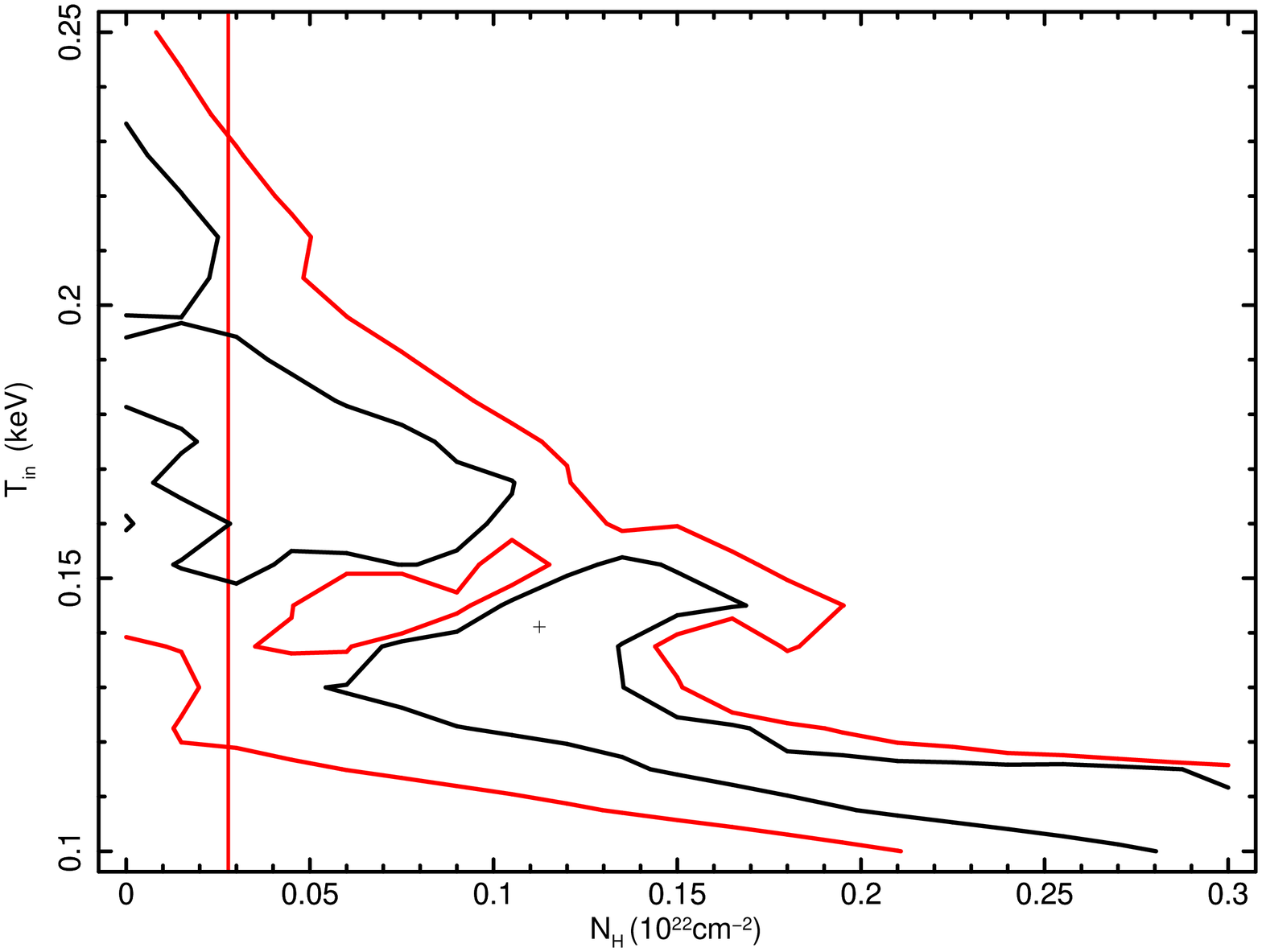}

	\end{minipage}
	\begin{minipage}{0.45\linewidth}
	\centering

    \includegraphics[height=\linewidth,angle=-90]{f8b.ps}

	\end{minipage}
	\centering

    \caption{1$\sigma$ and 2$\sigma$ contours for the two-component PO+DBB modeling of observations 2\&3 of S102. Left indicates values of \NH\ and \kt\ and the right panel presents the $\Gamma$ and \kt\ values from the same fit. The best-fit values are indicated by the cross hairs in both panels. In the left panel the solid vertical line indicates Galactic \NH. \label{fig:Src1022con} }

\end{figure}

The best-fit spectra of these four sources are all summarized in Table \ref{tab:Bestfit} and the confidence contours of S74 are presented in Figure \ref{fig:Src74con}. The confidence contours from the single-component PO model of the three spectral groupings from S102 are presented Figure \ref{fig:Src102con} and the contours from the two-component model of observations 2 \& 3 are shown in Figure \ref{fig:Src1022con}.

\section{Simulations of LMXB Spectra}
\label{sec:sims}

From studies of Galactic binaries it has been observed that the energy spectra of LMXBs often exhibit a composite shape of both thermal and non-thermal emission components. In these sources the thermal component is well modeled by the multicolor disc blackbody model, which originates in the inner accretion disc, and the non-thermal component is well described by a power law component. This composite emission provides a wide range in X-ray properties that can be classified into five distinct spectral states. In RM06 these states are defined to be; the quiescent state, a non-thermal hard state with very low emission (\LX\ $10^{30.5}$-$10^{33.5}$ \ergps), the hard state (denoted hereon in as NT; non-thermal state, to avoid confusion with the high/soft state classification that can also be used to describe LMXB spectral states), where the non-thermal component dominates the spectra ($>$80\%\footnote{Where the full-band flux is 2$-$20 keV from {\em RXTE} spectra, not the 0.3$-$8.0 keV band that has been used in the rest of this paper. Although, definitions from RM06 do provide a fairly accurate interpretation of spectral states observed in the \CHANDRA\ bandpass.}) and the spectral index is in the range 1.5$<\Gamma<2.1$, and the thermally dominant state (TD), where the thermal emission arising from the disc contributes to more than 75\% of the flux. In the TD state the disc temperature is typically in the range of 0.7$-$1.5 keV and the faint power law component is steep ($\Gamma>$2.1). Sources have also been observed in the steep power law (SPL) state, where the sources are much brighter (\LX$>$0.2\Ledd) with a significant flux contribution arising from the non-thermal component with steeper slope of $>$2.4, compared to $\sim$1.7 in the hard state. The final spectral class encompasses emission not defined by the other states, where spectra appear to show sources between states, this is termed the intermediate state. It should be noted that this definition does not represent a single state but indicates that the source emission cannot be described by any of the four previous source classes.

From \CHANDRA\ and \XMM\ observations of extra-galactic LMXBs there are typically too few counts to apply composite models to each spectrum, therefore single component models that provide an approximation of the spectra must be applied to the data. To understand how well these single-component models describe the X-ray emission of LMXBs, simulations have been performed with Sherpa (v3.4), using \CHANDRA\ response files. In these simulations spectra have been produced from two-component models comprising both a MCD (XSDISKBB in Sherpa; the XSPEC DBB model used in section \ref{sec:results}) and a PO component, along with the $phabs$ photoelectric absorption model which has a value of Galactic absorption. This model covers the 0.3$-$8.0 keV \CHANDRA\ energy range. These generated spectra have then been fitted to single-component models to characterize the typical best-fit values that are recovered. To broadly cover physical properties that have been observed in the spectra of Galactic LMXBs (RM06) simulations were run over a large parameter space where the inner disc temperature of the DBB model ranged from 0.50 keV to 2.25 keV in steps of 0.25 keV and input $\Gamma$ slopes increased in steps of 0.2, from 1.5 to 2.5. Further, to investigate how the composite spectral shape of sources in hard, thermally dominant and steep power law states were represented by single-component fits, different flux ratios of 75\% ,60\%, 50\%, 40\%, 25\%, 10\% and 0\% were also included in the simulations, where the percentage indicates the flux contribution from the thermal component. This was also expanded to include a sub-set of parameters at 90\% and 100\%, where spectra with $\Gamma$ fixed at 1.7 and 2.5 were produced for the full disc temperature range. It was necessary to produce only a sub-set of these flux ratio values due to the computer intensive nature of these simulations. A value of 2.5 was selected for $\Gamma$ based on typical LMXB photon index values of sources in a thermally dominant state. A photon index of 1.7 was selected based on the results from the lower flux ratio simulations, where this input parameter was found to provide median values of \NH, $\Gamma$ and \kt\ for the single component best-fit parameters over the full photon index range.

To replicate the quality of data that has been presented in this paper, the generated models were scaled to produce spectra of 1000 counts, with subsequent runs also covering a sub-set of 500 and 250 count data. After a spectrum was generated in Sherpa, using the {\em fakeit}\footnote{http://asc.harvard.edu/sherpa3.4/threads/fakeit/} command, single-component PO and DBB models were fitted and the best-fit values recorded. Following this, the original two-component model was fitted to the data to ascertain how reliably the low-count spectrum could recover the input parameters. For each set of parameters 100 spectra were generated, providing a measure of the standard deviation of the best-fit values. 


\subsection{Single Component Spectra}
\label{subsec:single}

\subsubsection{Single PO Fits}
\label{subsubsec:po}

\begin{figure}
  \begin{minipage}{0.485\linewidth}
  \centering
  
    \includegraphics[width=\linewidth]{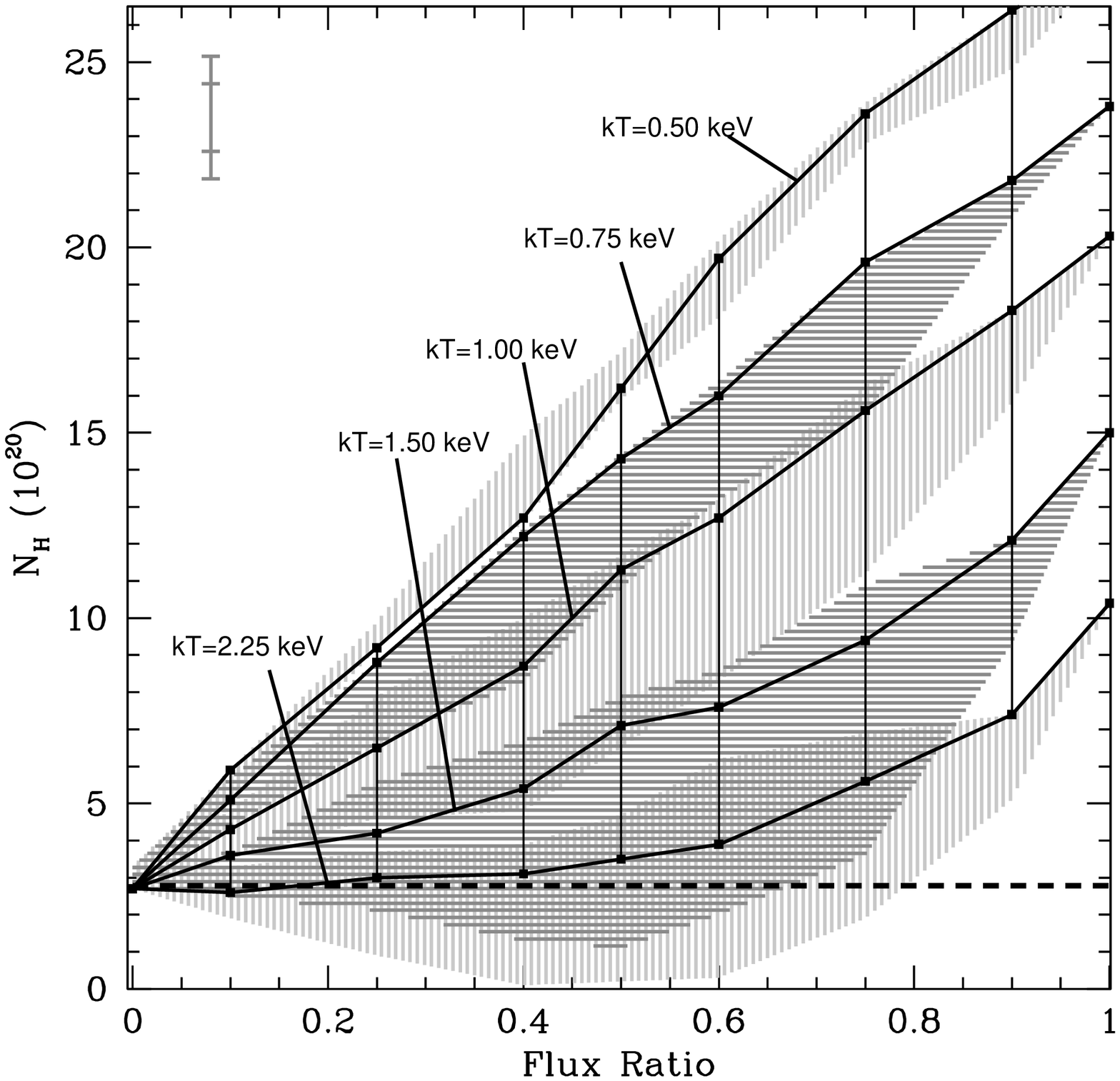}
  
  \end{minipage}\hspace{0.02\linewidth}
  \begin{minipage}{0.485\linewidth}
  \centering

    \includegraphics[width=\linewidth]{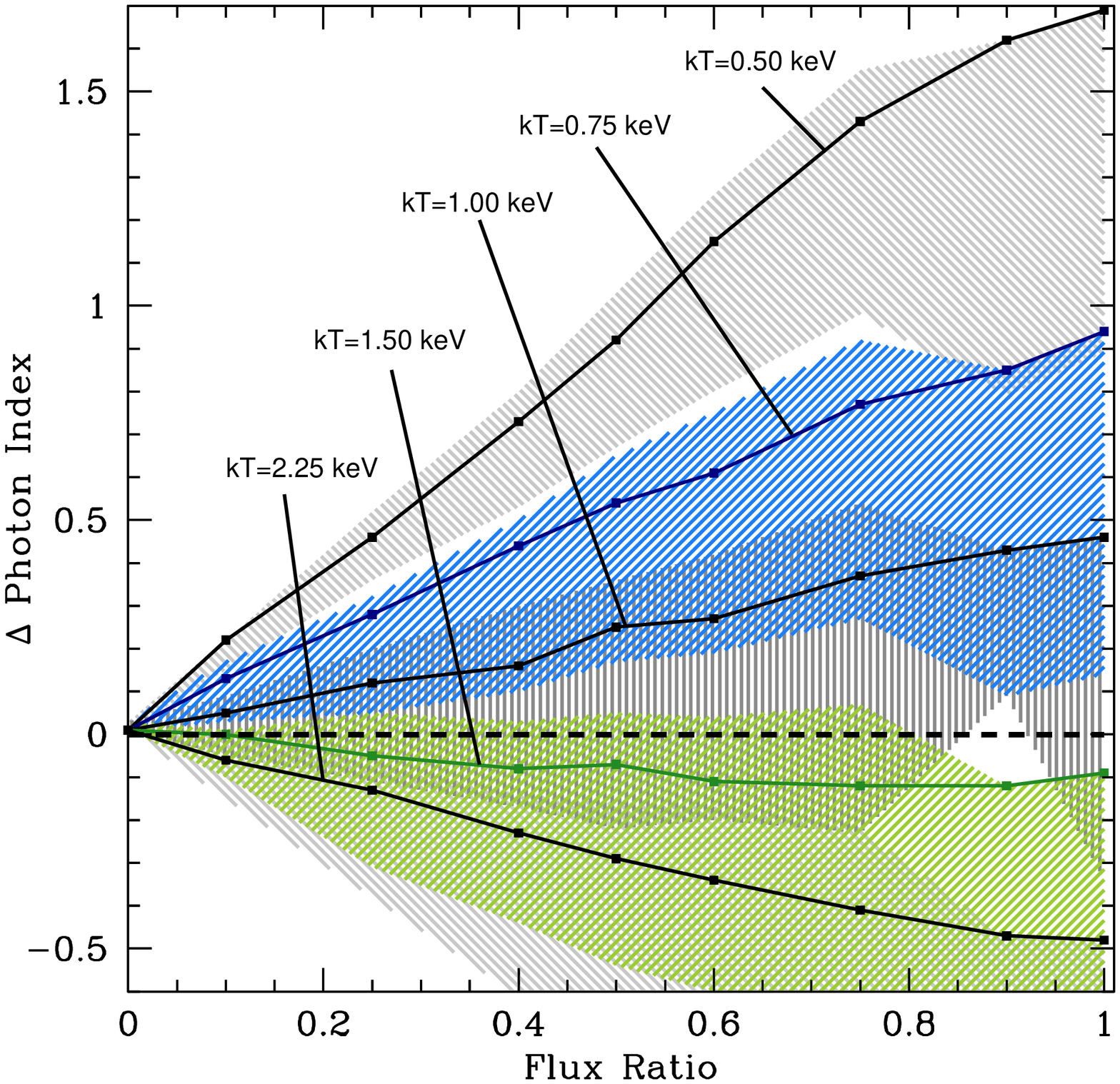}

\end{minipage}
  
  \centering
  \caption{Results of the best-fit values derived from the single-component PO model applied to the 1000 count simulated spectra (detailed in section \ref{sec:sims}), where all spectra were simulated with Galactic \NH. The left panel indicates the values of \NH\ plotted against the flux ratio of the disc-to-total flux. The right panel shows the difference ($\Delta$) between the input and best-fit value of $\Gamma$. In both plots the solid diagonal lines label input $\Gamma=1.7$ for each simulation, for input \kt\ values of 0.50, 0.75, 1.00, 1.50, and 2.25 keV. The shading indicates the spread in best-fit values for the full range of photon indices 1.5$-$2.5. In the left panel the horizontal dashed line indicates the Galactic \NH\ an and the vertical solid lines are grid line. The error bars in the upper-left corner indicate the standard deviation of the 0.75 (large) and 0.25 (small) value of \NH. In the right panel the \kt\ values are presented in different colors for clarity. It should be noted that different shading patterns of green, blue and gray are a consequence of overlapping shading and do not indicate additional information within the figure. \label{fig:simspo} }

\end{figure}

In Figure \ref{fig:simspo} a summary of the results of fitting the single-component PO model is presented. In the left-hand panel the best-fit value of \NH\ for each of the parameter models is shown and in the right-hand panel the difference ($\Delta$) between input and best-fit photon index is presented. In both panels the best-fit values are plotted against increasing thermal flux ratios, with the solid diagonal lines indicating the best-fit values for $\Gamma$=1.7, for each temperature step of the disc component (a sub-set of temperatures are shown for clarity in the figure). The shaded regions indicate the maximum and minimum value of \NH\ over the full photon index range for each temperature step (where only photon index values of 1.7 \& 2.5 have been simulated for flux ratio values of 90\% and 100\%). In the left-hand panel the horizontal dashed line indicates the value of Galactic \NH\ for NGC 3379. The standard deviation for the 1000 count spectra for the 75\% and 25\% flux ratios are indicated by the error bar in the top left corner of the plot, where the smaller error bar indicates the 25\% flux ratio standard deviation ($\sigma$=1.82$\times10^{20}$cm$^{-2}$) and the larger error bar the standard deviation from the 75\% flux ratio model ($\sigma$=3.30$\times10^{20}$cm$^{-2}$).

From these figures it is clear that as the spectra become more disc-dominated the best-fit values from the single component PO model increasingly diverge from the input parameters of the two-component model. Importantly, for all values of \kt\ and $\Gamma$, best-fit values of \NH\ are above Galactic (excepting the spectra generated from the extreme models with input parameters $\Gamma$=2.3$-$2.5 and \kt$\ge$1.50 keV) and even up to $\sim$8 times that of Galactic when the spectra are in a thermally-dominant state (i.e. the disc fraction contributes more than 75\% of the total flux). In fact, assuming a typical disc temperature of 1 keV, even with a disc contribution of only 40\% the recovered \NH\ from the single-component PO fit is $>$3 times Galactic \NH\ (a value significantly above the Galactic value even after considering the standard deviation). Further, unsurprisingly, the difference between the input and best-fit photon index (|$\Delta\Gamma$|) increases as the ratio of disc flux to total flux increases, indicating that, as the non-thermal component becomes less prominent within the composite fit, the reliability of the $\Gamma$ recovered from the single component model is reduced.

\begin{figure}
  \centering
    \includegraphics[width=0.6\linewidth]{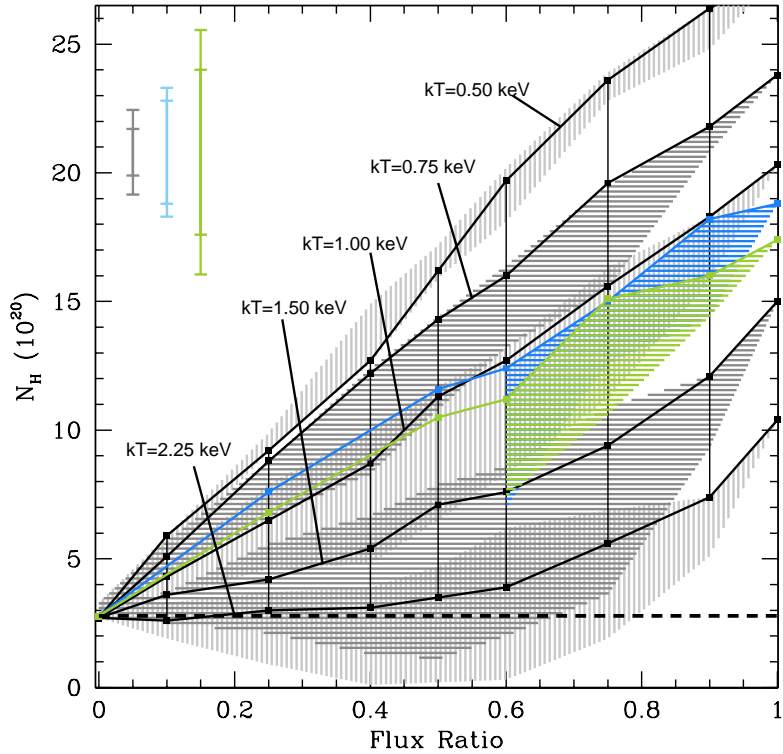}
    \caption{This figure presents the same information as is shown in the left panel of Figure \ref{fig:simspo} but also includes the best-fit values of \NH\ derived from lower count simulations of input parameters $\Gamma$=1.7 \& 2.5 and \kt=1.00. The 500 count data is plotted in blue and the 250 count data in green. The results of the $\Gamma$=1.7 simulations are labeled by the solid diagonal lines and the best-fit values of \NH\ from the $\Gamma$=2.5 simulations for 60\%, 75\%, 90\%, and 100\% disc-to-total flux ratios, are indicated by the outer edge of the shading (follwing the findings of the 1000 counts simulations, intermediate values of $\Gamma$ are expected to reside within the shaded region). Error bars in the top left indicate the standard deviation for 0.75 (large) and 0.25 (small) flux ratios, these are also color-coded. The horizontal dashed line indicates Galactic \NH. \label{fig:simslowC} }

\end{figure}

A sub-set of these simulations were also generated for spectra with 500 and 250 counts, where flux ratios models of 100\%, 90\%, 75\%, 60\%, 50\%, 25\% and 0\%, were produced with an inner disc temperature of 1 keV and photon index of 1.7. In addition, for the higher flux ratio values $\ge$60\%, models with $\Gamma$=2.5 were also used as input parameters for the simulations to represent thermally dominant states. The best-fit values of \NH\ with lower count spectra are compared with the full 1000 count simulations in Figure \ref{fig:simslowC}, where the 1000 count spectral results are presented in an identical manner to the left-hand panel Figure \ref{fig:simspo}. The best-fit \NH\ values of the 500 and 250 count spectra are indicated by the blue and green lines respectively, with the shading indicating the best-fit values from the $\Gamma$=2.5 simulations. As in Figure \ref{fig:simspo}, the standard deviation of the 75\% and 25\% flux ratios are indicated by the color-coded error bars in the top left corner of the plot. From this figure it is clear that, despite the larger standard deviations in the lower-count models, the best-fit values of \NH\ are almost identical to those obtained with the 1000 count data. Even in the case of 250 counts, when a LMXB is in a 'typical' thermal-dominant state (i.e. \kt=1 keV, $\Gamma$=2.5 and disc flux $>$75\%) the best-fit value of \NH\ is still more than three times that of Galactic absorption.

\subsubsection{Single DBB Fits}
\label{subsubsec:dbb}

\begin{figure}
\begin{minipage}{0.485\linewidth}
  \centering

    \includegraphics[width=\linewidth]{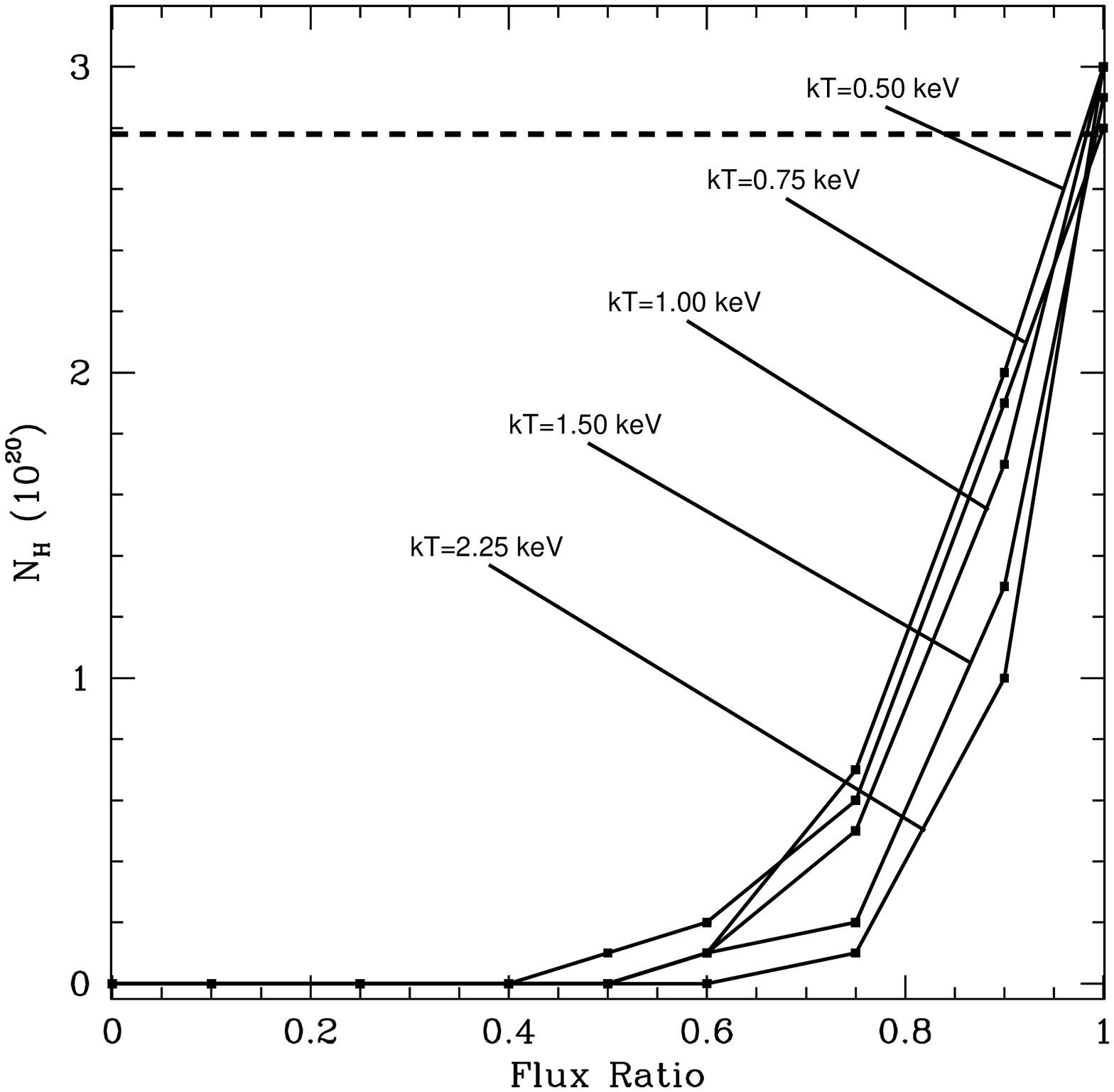}

 \end{minipage}\hspace{0.02\linewidth}
\begin{minipage}{0.485\linewidth}
  \centering
  
    \includegraphics[width=\linewidth]{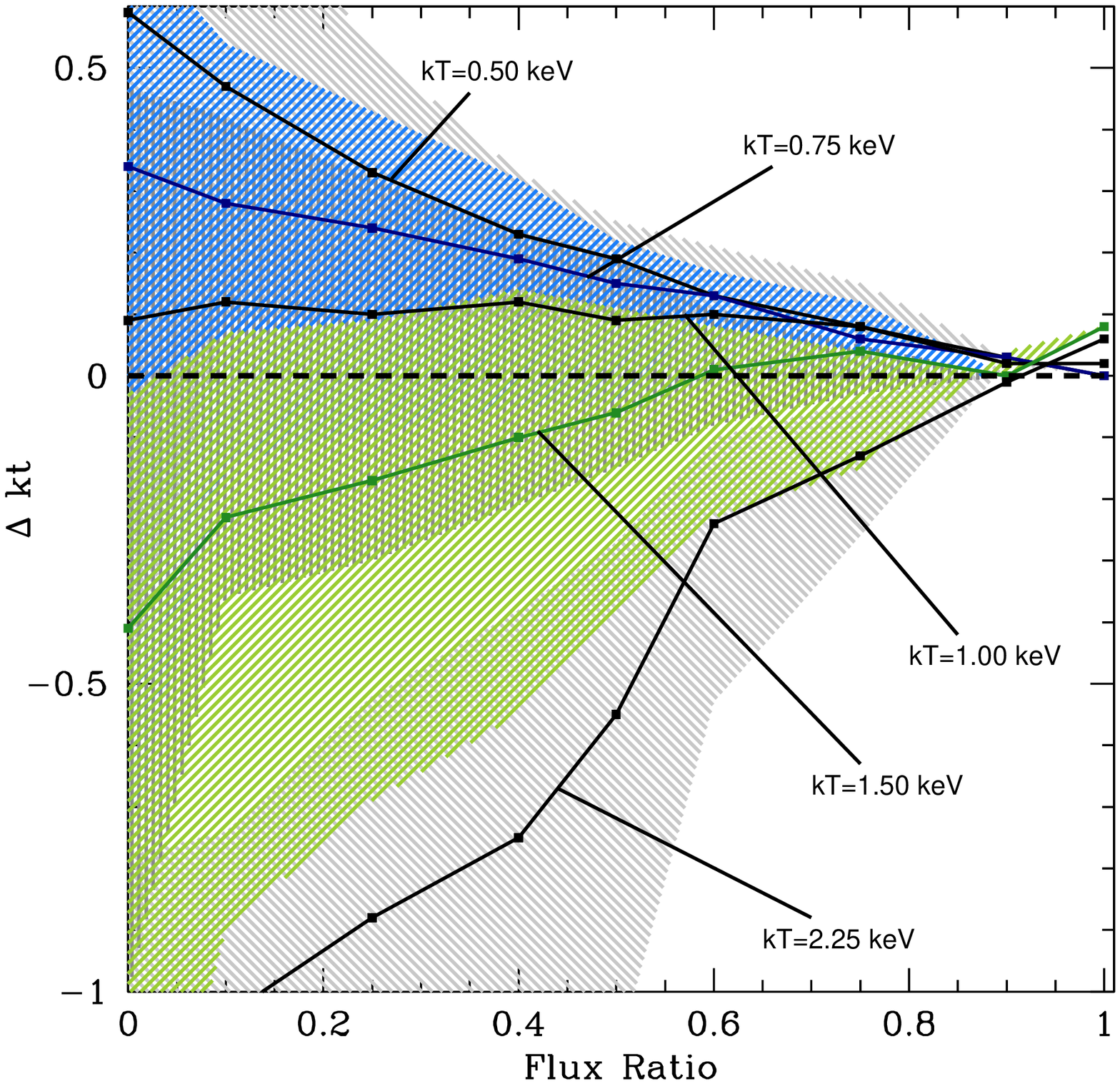}
  
  \end{minipage}\hspace{0.02\linewidth}

  \centering
  \caption{This figure presents the best-fit values derived from the single-component DBB model applied to the simulated spectra with input values of \NH\ set to Galactic absorption (detailed in section \ref{sec:sims}). As in Figure \ref{fig:simspo}, the left panel indicates the values of \NH\ plotted against the flux ratio of the disc-to-total flux and the right panel shows the difference ($\Delta$) between the input and best-fit value of \kt. In both plots the solid diagonal lines labels input $\Gamma=1.7$ for each simulation, for input \kt\ values of 0.50, 0.75, 1.00, 1.50, and 2.25 keV. Shading is not presented in the left panel due to the small variations of output \NH\ values but in the right panel it indicates the spread in best-fit values for the full range of photon indices 1.5$-$2.5. It should be noted that different shading patterns of green, blue and gray are a consequence of overlapping shading and do not indicate additional information within the figure. The Galactic absorption column is indicated in the left panel by the horizontal dashed line.  \label{fig:simsMCD} }

\end{figure}

A summary of fitting the single-component multicolor disc models from the simulations is presented in Figure \ref{fig:simsMCD}. In the left-hand plot of this figure the \NH\ values from the DBB  model with a fixed PO component of $\Gamma$=1.7 are plotted for disc temperatures 0.50 keV, 0.75 keV, 1.00 keV, 1.50 keV and 2.25 keV. No shading is shown for the different photon index slopes in this panel due to the small deviations in best-fit values of \NH. Galactic \NH\ is indicated by the dashed horizontal line. In the right-hand panel $\Delta$\kt\ is shown against increasing flux-ratios, where the variation of best-fit \kt\ for different values of photon index is indicated by the shaded regions. The $\Gamma$=1.7 input parameter values are shown by the solid lines.

From these figures it can be determined that when the flux emission does not arise entirely from the disc component of the LMXB, the best-fit value of \NH\ from the single component DBB model is $<$ Galactic absorption. In fact, even with a thermal flux ratio of 75\% the best-fit value of absorption is more than 3 times smaller than that of Galactic, tending to 0 as the flux ratio decreases. In the right-hand panel it is shown that as the input parameters of the simulations generate lower flux ratio spectra, so the value of |$\Delta$\kt| increases. This indicates that as the composite model becomes more dominated by the non-thermal component the best-fit parameters from the DBB model become less reliable, as was seen with the single-component PO model for the disc-dominant spectra.

\subsubsection{Composite PO+DBB Fits}
\label{subsubsec:dbbpo}

In addition to the single component fits, a composite PO and DBB model, with initial parameters set to the input values of the simulation, was also applied to each of the spectra. In all cases the fit was statistically acceptable, however, from comparing the best-fit values to the input parameters the two component model was unable to adequately recover the initial values. This is due to the degenerate nature of the disc and PO components of the model and indicates that, in general, for sources with less than 1000 counts a two component model will be unable to provide well constrained parameters (although this will be dependant on the shape of the spectrum).

\subsubsection{\NH\ as a Discriminant}
\label{subsubsec:nh}

These simulations provide a robust framework in which low-count single-component spectra can be used to infer properties of LMXBs and their spectral state. From these spectra it has been demonstrated that, whilst PO or DBB models are not capable of accurately constraining the spectral parameters of X-ray binaries when both a strong thermal and non-thermal component are present in the spectrum, they can provide important information that can be used to aid the interpretation of the source's spectral state. From the single-component PO model the \NH\ parameter is a strong indicator of the flux ratio between the two spectral components, where an elevated value of intrinsic absorption is indicative of the source containing a significant disc component (of at least 25\% flux for lower inner disc temperatures). In fact, this thermal to non-thermal contribution can be further constrained by determining the \NH\ absorption value from the DBB fit, where only sources in a thermally-dominant state provide values that are consistent with Galactic \NH. Further, these simulations have also determined the reliability of the photon index or \kt\ values recovered from the single-component models for these composite sources. In the case of the PO model, when the source is in a hard state, indicated by a value consistent with Galactic \NH, the best-fit value of $\Gamma$ provides an accurate value of the input parameter. Similarly, when the best-fit value of \NH\ from the DBB model provides values $>$ 0, indicating the source is in a thermally dominant state, the best-fit \kt\ provides a reliable value of the true inner disc temperature. In all other instances neither models provide a reliable measure of the intrinsic source properties.

\subsubsection{Flux Determination}
\label{subsubsec:flux}

\begin{figure}
\begin{minipage}{0.485\linewidth}
  \centering

    \includegraphics[width=\linewidth]{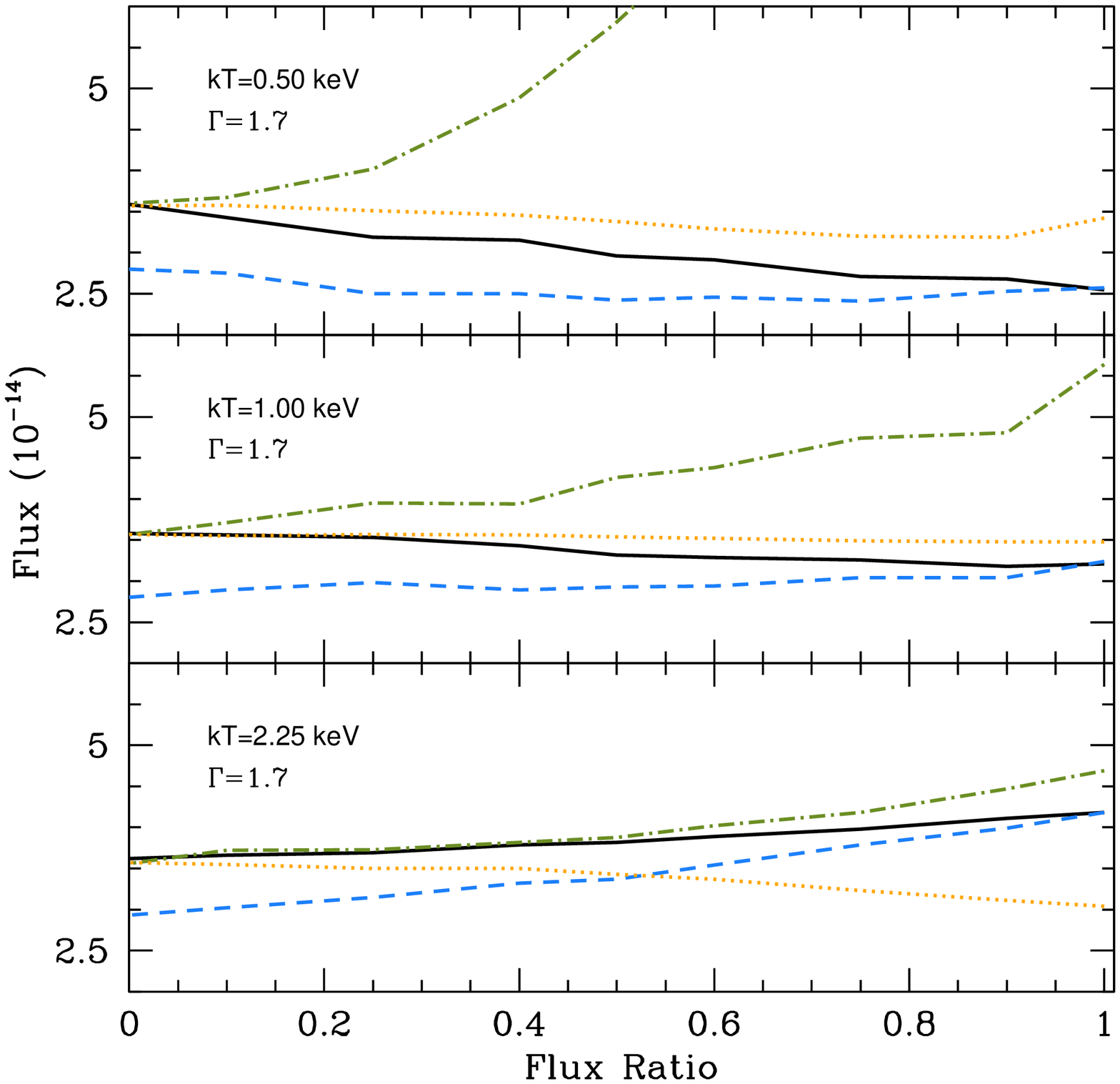}

 \end{minipage}\hspace{0.02\linewidth}
\begin{minipage}{0.485\linewidth}
  \centering
  
    \includegraphics[width=\linewidth]{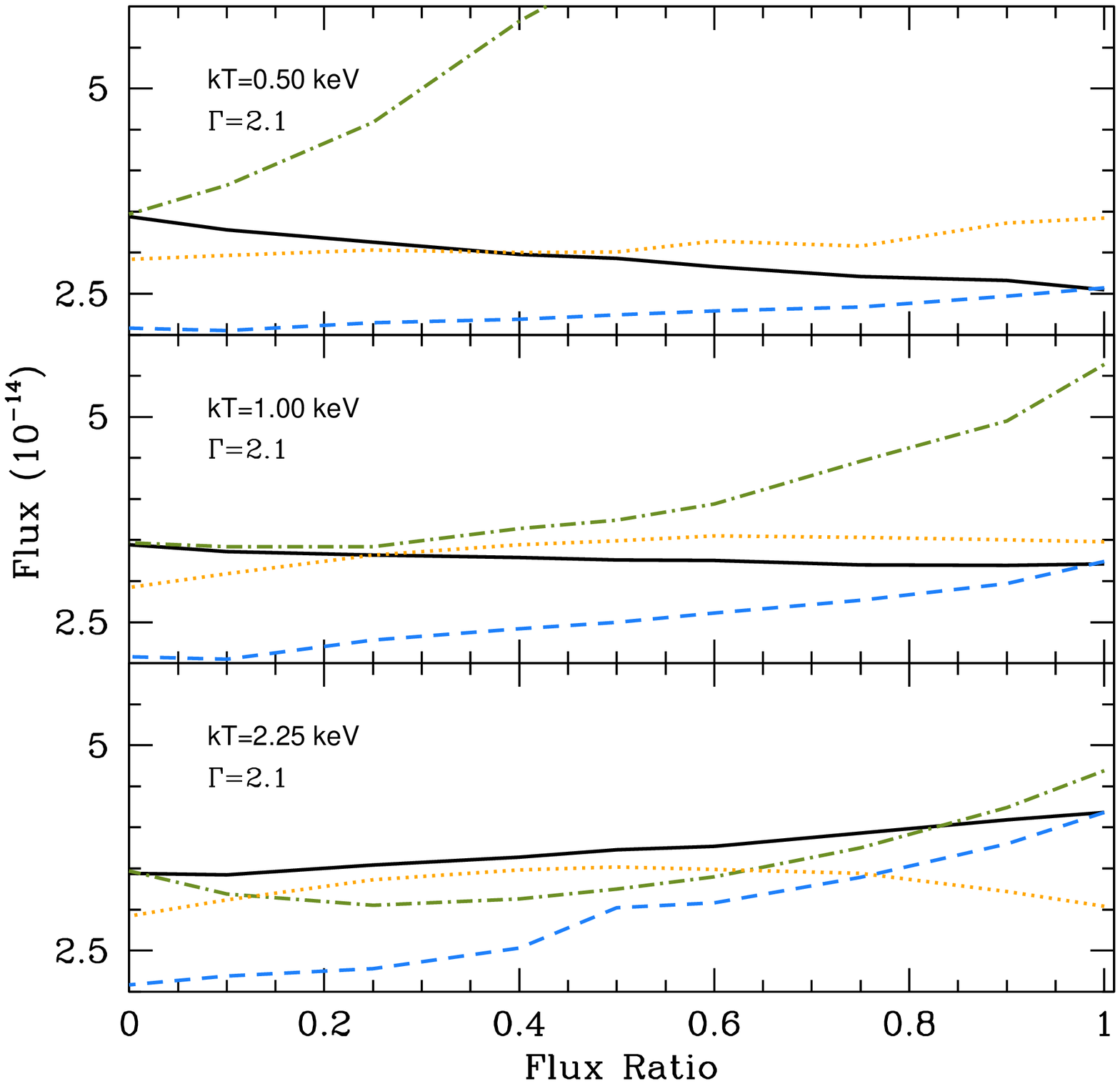}
  
  \end{minipage}\hspace{0.02\linewidth}

  \centering

    \caption{\LX\ plotted against the flux ratio, where the 0.3$-$8.0 keV flux from the input two-component PO+DBB model (black line) is compared to the \LX\ values derived from the single-component fits; the PO model is indicated by the dashed-dot green line and the DBB model by the dashed blue line. Flux derived from a canonical PO model of Galactic \NH\ and $\Gamma$=1.7 is indicated by the dotted orange line. The panels on the left indicate the simulations with an input $\Gamma$=1.7 and the right, the $\Gamma=2.1$ simulations. Input \kt\ values are labeled in each panel, as is the input photon value. \label{fig:simslx} }

\end{figure}

In addition to the $\Gamma$, \kt\ and \NH\ parameters, 0.3$-$8.0 keV values of flux from the single component models were also compared to the input flux values. A comparison of the derived fluxes for disc temperatures of 0.50, 1.00 and 2.25 keV, and photon index values 1.7 and 2.5 are shown in Figure \ref{fig:simslx}. In this figure the fluxes from the three different models, as well as a canonical PO model of Galactic \NH\ and $\Gamma$=1.7, are show in six panels, where the left-hand panels indicate fluxes for all values of \kt\ from the $\Gamma$=1.7 models, and the right-hand panels presents the flux values from the $\Gamma$=2.5 models. The fluxes derived from the composite models are indicated by the solid black lines, the single PO model by the dot-dashed green lines, the best-fit values from the DBB model the dashed blue lines and the canonical model flux is indicated by the dotted orange line. 

From this figure it can be seen that as the disc component becomes more prominent in the composite model, the value of flux derived from the single-component PO model diverges from the flux determined from the composite model.  The value of this divergence increases as the temperature of the inner disc decreases. Further, by comparing these changes in flux with Figure \ref{fig:simspo} it can be seen that in all instance where an elevated value of \NH\ is determined from the PO model, the intrinsic flux from the fit is significantly higher than the true source value, a consequence of the excess soft emission of the disc being modeled by the higher absorption component. 
Conversely, the intrinsic flux from the DBB best-fit model show a lower value than the one defined by the input model in all instances, up to a disc contribution of 100\%. Although, these lower values provide a much closer value to the true flux than that provided by the single-component PO fit (for instances where absorption column values above Galactic \NH\ are recovered). 
However, the flux values that are closest to the input model values are derived from the canonical PO model. This is true in most cases where the fitted \NH\ from the single PO model is significantly above the Galactic value, for both input $\Gamma$=1.7 and 2.5. However, this canonical PO model does provide a less accurate value of the true flux compared to the DBB model when a significant disc is present. 
Therefore, when providing flux values for sources with parameters determined from single-component models (when \NH\ is enhanced), the flux from the canonical model should be used as an estimate of the true flux, except when the DBB model best-fit value of \NH\ is consistent with Galactic absorption, indicating that the source is in a TD state. In these instances the DBB flux value should be taken as a closer estimate of the source flux.

It should be noted that the interpretation of these spectral models has made the assumption that these sources are not embedded within diffuse emission arising from the hot gas within the galaxy. In the instances where the galaxy contains a substantial ISM care should be take to determine if the enhanced values of \NH\ are a consequence of intrinsic emission or caused by the surrounding medium of the source. Further, intrinsic absorption from the source itself, e.g. a warped accretion disc; as proposed for the BH-GC source in NGC 4472 by Shih \etal\ (2008), cannot be ruled out when enhanced values of \NH\ are determined in the spectra.

\subsection{Cool-Disc Models}
\label{subsec:cool}

In the previous section, simulations covering the range of parameters that have been observed in LMXBs in three spectral states: hard, thermally dominant and the steep power law state, have been presented. In this section the simulations are extended to include spectra that have been observed in X-ray binaries with SPL characteristics, but have cooler inner disc temperatures than are typically observed for sources in this state (\kt$<$0.4 keV). From spectral modeling of {\em RXTE} observations of Galactic sources with this emission (e.g.  XTE J1550$-$564, Kubota \& Done 2004) it has been suggested that these sources appear to have a cooler disc component due to the presence of a Comptonizing corona coupling with the disc, altering the intrinsic properties of the accretion disc (Done \& Kubota 2006). This type of cool-disc spectrum has also been observed in ULXs, albeit with shallower photon indices ($\sim$1.7, e.g. Soria \etal\ 2007). This ULX emission has been likened to the spectra observed in Galactic binaries in outburst (Kubota \& Done 2004), where the explanation of a Comptonizing component altering the properties of the accretion disc has again been suggested.

\begin{figure}
\begin{minipage}{0.485\linewidth}
  \centering

    \includegraphics[width=\linewidth]{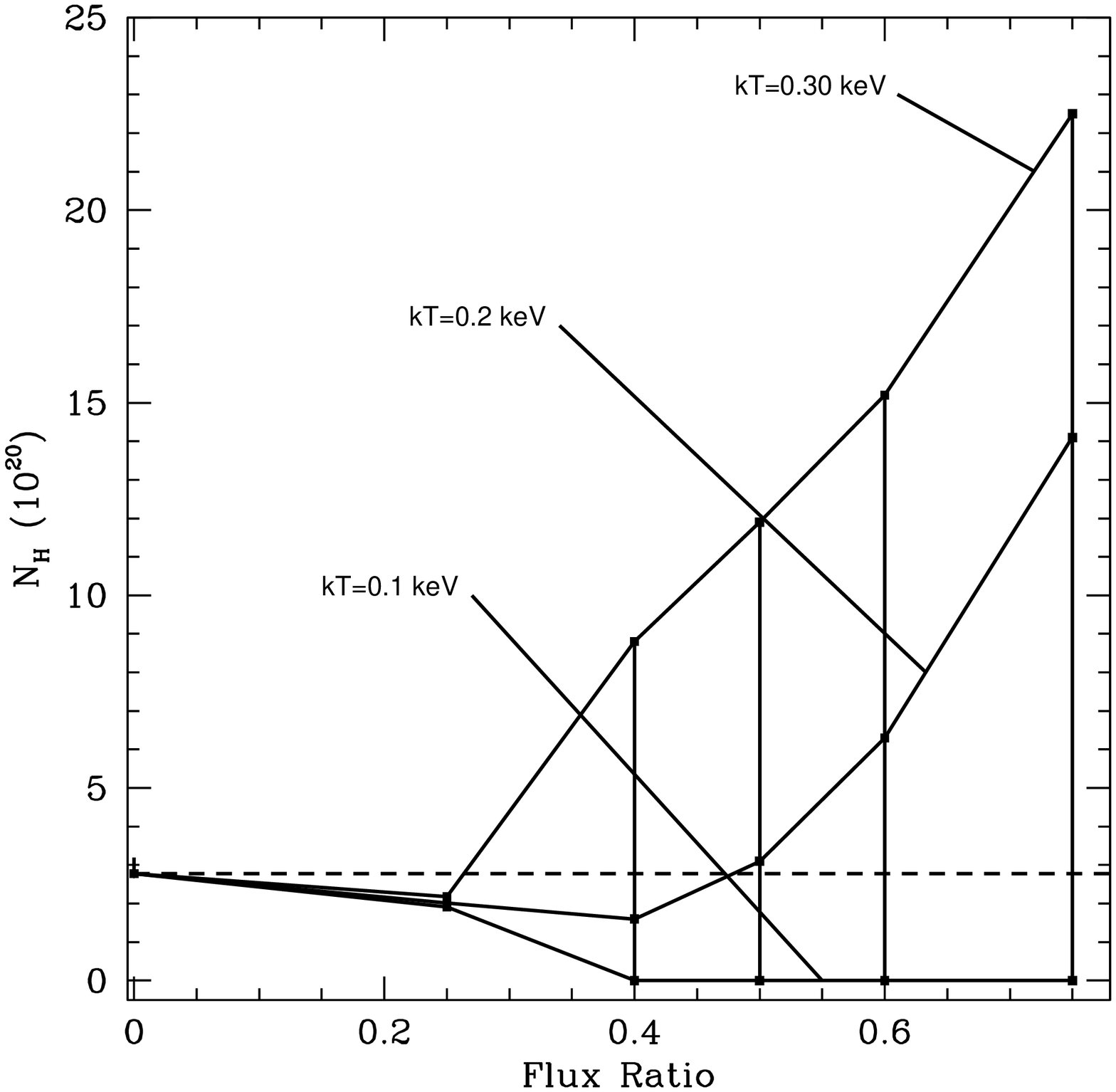}

 \end{minipage}\hspace{0.02\linewidth}
\begin{minipage}{0.485\linewidth}
  \centering
  
    \includegraphics[width=\linewidth]{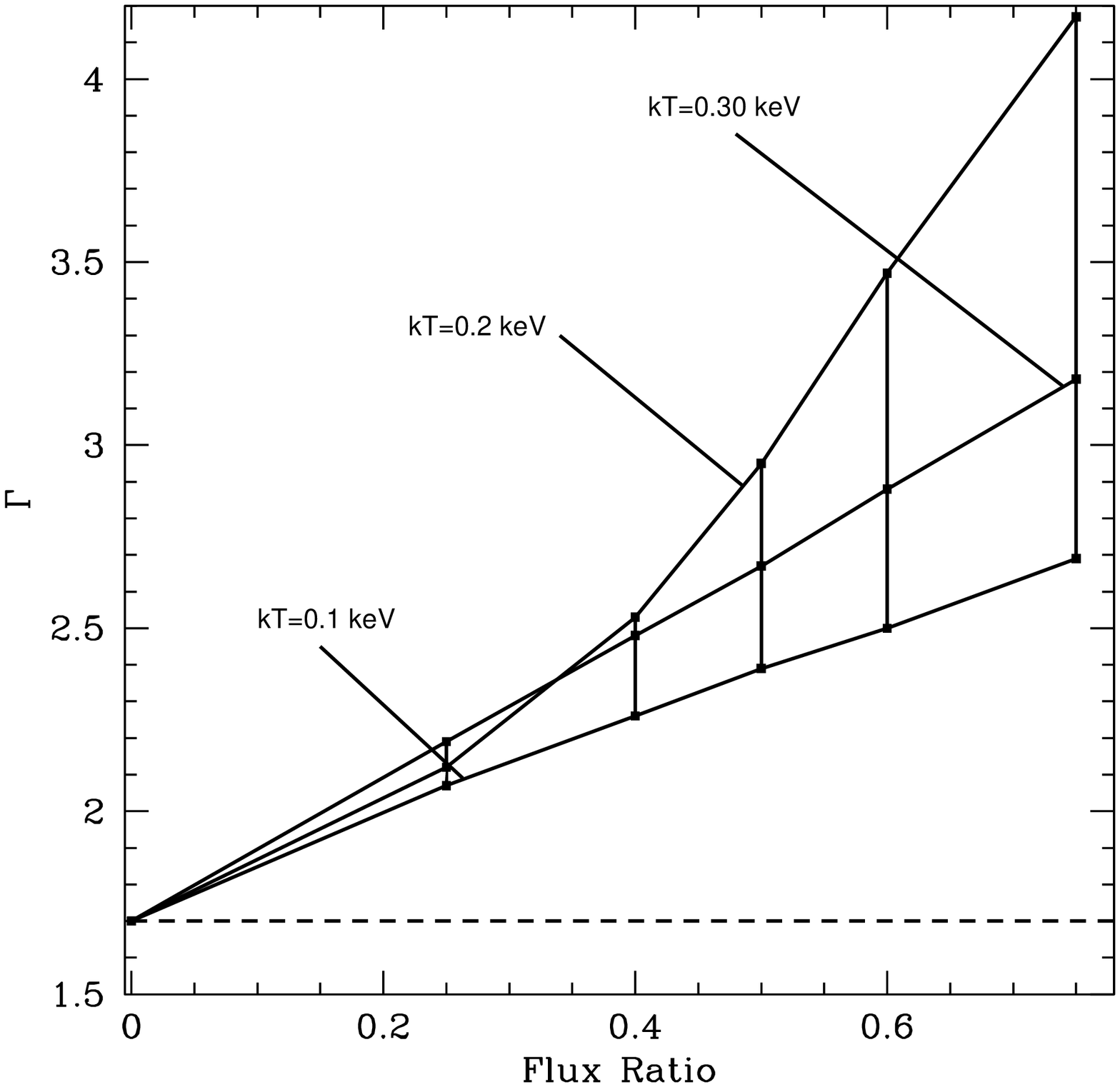}
  
  \end{minipage}\hspace{0.02\linewidth}

  \centering
  \caption{Best-fit results from the cool-disc simulations presented in section \ref{subsec:cool}, following the same presentation as Figure \ref{fig:simspo}. Only the best-fit values derived from the $\Gamma$=1.7 simulations are presented. In the right panel the best-fit value of $\Gamma$ is shown (instead of $\Delta$ photon index). \label{fig:simscool} }

\end{figure}

To investigate the range of parameters obtained from single component modeling of this cool-disc spectral state, simulations were produced with input inner disc temperatures of 0.1, 0.2 and 0.3 keV, and photon index values of 1.7 and 2.5; representing both ULX and Galactic LMXB spectra. Flux ratio values of 75\%, 60\%, 50\%, 40\%, 25\% \& 0\% were produced for each set of input parameters. A summary of the single component PO models of these simulated spectra are presented in Figure \ref{fig:simscool}, following the same convention as in Figure \ref{fig:simspo}. In this figure only the $\Gamma$=1.7 input models are indicated in both plots for clarity. From the left-hand panel in this figure, where values of \NH\ are presented, it can be seen that when the source has a high flux ratio (with \kt$>$0.1 keV) the best-fit value of \NH\ is much higher than that of Galactic, as was seen in the previous simulations. As the non-thermal component becomes more prominent the value of \NH\ decreases and in fact, for flux ratio values $<$25\%, the best-fit value of \NH\ is either consistent with or below Galactic for disc temperatures of 0.3 keV, and below Galactic with a flux-ratio value of $<$50\% for \kt=0.2 keV. For the simuations with an inner disc temperature of 0.1 keV all values of \NH\ are 0, only becoming consistent with Galactic \NH\ with flux ratios $<$25\%. 

The best-fit values of \NH\ derived from the simulations with the non-thermal component described by a PO with $\Gamma$=2.5, exhibited similar variations to those presented in Figure \ref{fig:simscool}, with all \NH\ values from the 0.1 keV inner disc simulations $<$ Galactic \NH. However, there were differences between the 0.2 and 0.3 keV simulations from the different $\Gamma$ values, with the single component fits from the steeper value of 2.5 always determining a best-fit \NH\ higher than Galactic for all flux ratio values, apart from 0\%.

The single-component DBB models for the cool-disc simulations were statistically rejected in 100\% of cases when the input photon index was 1.7. When a slope of 2.5 was used, the 0.2 and 0.3 keV disc temperature simulations were adequately described by single-component models in $<$50\% of cases. When these fits were statistically acceptable the best-fit value of \NH\ was 0. When the inner-disc temperature was 0.1 keV, none of the simulated spectra were statistically acceptable. 

\subsection{Different Absorption Models}
\label{subsec:absorp}

In the simulations that have been presented in this paper the $phabs$ model (Baluci\'nska-Church \ McCammon 1992) has been used to describe the absorption of X-rays in the interstellar medium. However, other models of neutral absorption such as $wabs$ (Morrison \& McCammon 1983) and $tbabs$ (Wilms \etal\ 2000) are also commonly used to describe the emission from X-ray binaries.

\begin{figure}
  \centering
    \includegraphics[width=0.6\linewidth]{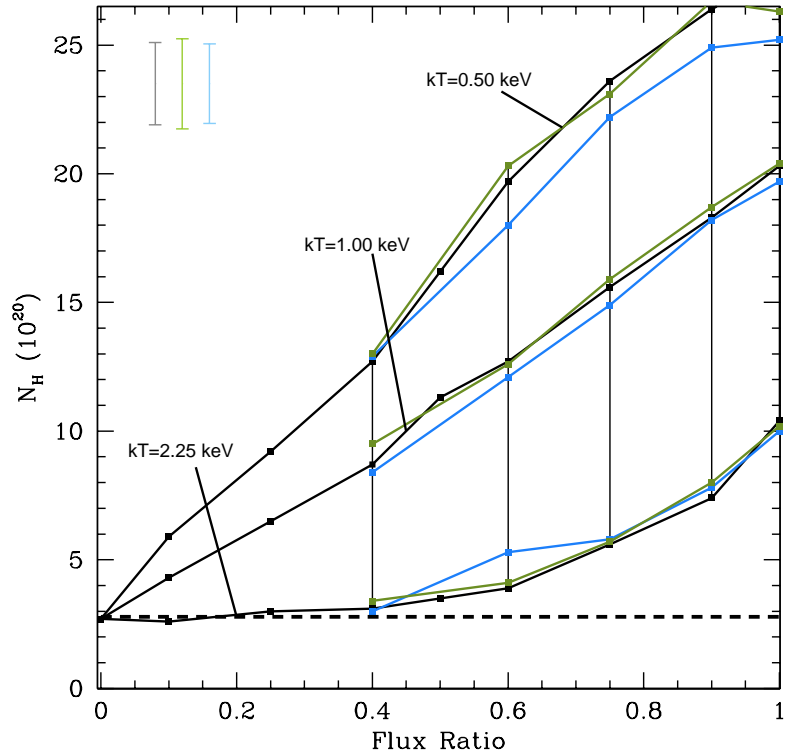}
    \caption{Best-fit results of simulations using three different photoelectric absorption models (detailed in section \ref{subsec:absorp}) following the same presentation as in the left-hand panel in Figure \ref{fig:simspo}. Here, only the best-fit values derived from the $\Gamma$=1.7 simulations are presented for values of \kt=0.50, 1.00 and 2.25 keV, over flux ratios values of 40\% to 100\%. The results from the {\em phabs} absorption simulations are labeled by the black diagonal lines, the results from the {\em tbabs} simulations are shown by the green lines and the results from the {\em wabs} simulations are plotted in blue. Error bars in the top left indicate the standard deviation for each set of simulations, these are also color-coded. The horizontal dashed line indicates Galactic \NH. \label{fig:simsabs} }

\end{figure}

To determine if the results from the $phabs$ simulations are consistent with results from alternative absorption models, a sub-set of simulations were generated using both $wabs$ and $tbabs$ models. To investigate the range of parameters obtained from single component modeling of these different absorption models, simulations of a 1000 counts were produced for sources with input inner disc temperatures of 0.50, 1.00 and 2.25 keV, and with a photon index value of 1.7. Flux ratio values of 100\%, 90\%, 75\%, 60\%, and 40\% were produced for each set of input parameters. A summary of the best-fit \NH\ values from the single component PO models of these simulated spectra are presented in Figure \ref{fig:simsabs}, following the same convention as in the left-hand panel of Figure \ref{fig:simspo}. In this figure the diagonal black lines indicate the $phabs$ model, the green lines the $tbabs$ model and the blue lines the $wabs$ model. The standard deviation of each derived value of \NH\ is indicated by the color-coded errorbar in the top left corner. From this it can be seen that all three models result in similar values of \NH, and are identical within the standard deviation. The best-fit values of \NH\ from the single-component DBB model were also compared, and all three models again resulted in almost identical best-fit values of \NH, with the fitted absorption column tending to 0 as the flux ratio decreased. These comparisons demonstrate that when spectrally modeling low-count data with single-component models, the selection of either $wabs, phabs$ or $tbabs$ in the spectral model will result in the same behavior outlined in section \ref{subsec:single}. 

\subsection{Simulations Applied to Other Galaxies}
\label{subsec:nh0}

To investigate how the single-component best-fit parameters altered when applying the models to galaxies with different values of absorption column, a sub-set of simulations were performed. For the first case, simulations were produced for NGC 4278, an elliptical galaxy with a line of sight absorption below that of the current simulations (1.76$\times10^{20}$cm$^{-2}$ compared to 2.78$\times10^{20}$cm$^{-2}$)\footnote{This galaxy is the subject of a companion paper Fabbiano \etal\ (2010).}. These simulations were performed using response files from spectral extraction of sources within the galaxy, and input parameters were set to cover inner-disc temperatures of 0.50, 1.00 and 2.25 keV, with a fixed photon index of 1.7. Flux ratio values from 40\% to 100\% were investigated, with all simulated spectra produced with 1000 counts to allow comparisons to be made with the models presented in section \ref{subsec:single}. 
This sub-set of simulations was then extended to determine the best-fit parameters for systems with Galactic absorption values of 5$\times10^{20}$cm$^{-2}$ and 1$\times10^{21}$cm$^{-2}$, with input parameters defined to cover the same $\Gamma$, \kt\ and flux ratio values as for the NGC 4278 simulations.

\begin{figure}
  \centering
    \includegraphics[width=0.6\linewidth]{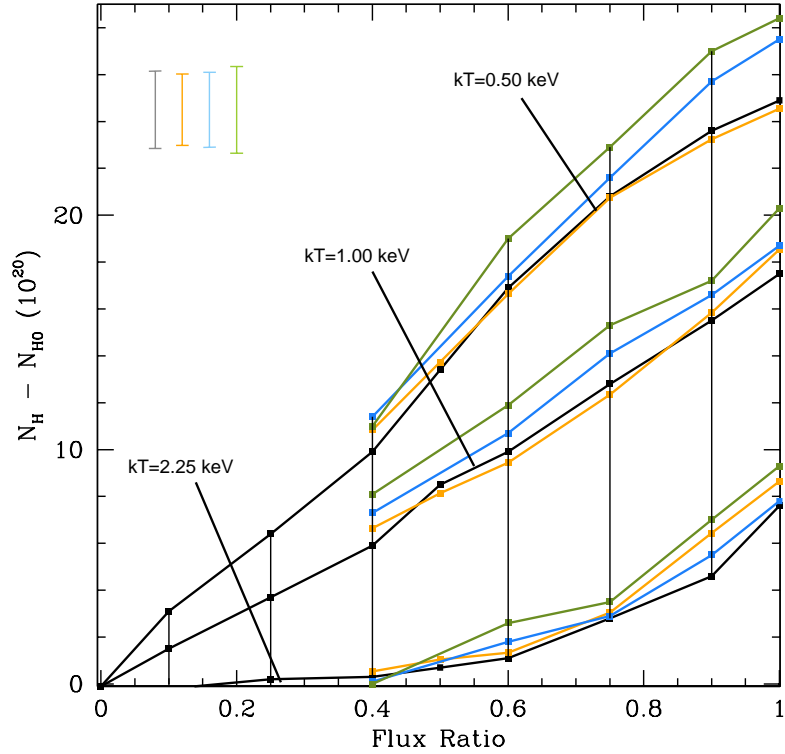}
    \caption{Best-fit results of simulations using four different Galactic absorption values (see section \ref{subsec:nh0} fore more details). In this figure, instead of presenting the best-fit value of \NH, as shown in Figure \ref{fig:simspo}, the value of excess \NH\ is plotted against the disc-to-total flux ratio, where the excess \NH\ is defined as the best-fit \NH\ value normalized by the Galactic absorption used in the simulations (\NH$-$\NH$_0$). Here, only best-fit values derived from the $\Gamma$=1.7 simulations are presented for values of \kt=0.50, 1.00 and 2.25 keV, over flux ratios values of 40\% to 100\%. The results from the main simulations are labeled by the black diagonal lines (with \NH$_0$=2.78$\times10^{20}$cm$^{-2}$), the results from the NGC 4278 simulations (\NH$_0$=1.76$\times10^{20}$cm$^{-2}$) are shown by the orange lines. The results from the \NH$_0$=5$\times10^{20}$cm$^{-2}$ simulations are plotted in blue and the \NH$_0$=1$\times10^{21}$cm$^{-2}$ simulations are indicated by the green diagonal lines. Error bars in the top left indicate the standard deviation for each set of simulations, these are also color-coded. \label{fig:sims4278} }

\end{figure}

A comparison of the results from these four sets of simulations is presented in Figure \ref{fig:sims4278}, where the excess absorption column values from the single-component PO models are shown. These excess \NH\ values are defined as the best-fit value of \NH\ normalized by the simulated Galactic absorption (\NH$_0$). In this figure the black lines indicate the simulations produced for NGC 3379, the orange lines the excess \NH\ values of the NGC 4278 simulations, blue the \NH$_0$=5$\times10^{20}$cm$^{-2}$ simulations and the green lines the \NH$_0$=1$\times10^{21}$cm$^{-2}$ simulations. In the top left corner color-coded error bars indicate the standard deviation of each set of simulations.
What can be seen from this figure is that all simulations exhibit the same behavior of increasing values of excess \NH\ as the thermal component of the source becomes more prominent. In fact, all excess absorption values from the four different sets of simulations are identical within the standard deviations. The spectra from these simulations were also fitted to single component DBB models, where the best-fit \NH\ values followed the behavior shown in Figure \ref{fig:simsMCD} .This indicates that these simulations are applicable in all galaxies, where single-component best-fit values can be used to determine the presence of absence of a thermal component, as well as the intrinsic source parameters.

\subsection{Decision Tree}
\label{subsec:tree}

Figure \ref{fig:flow} provides a summary indicating how the best-fit parameters derived from the single-component models can be used to determine the spectral state of a source. This is presented in the form of a decision tree, indicating how best-fit values of \NH\ from the PO and DBB models can be used to discriminate between NT, TD and intermediate states, including sources containing a cool-disc component.

\section{Discussion}
\label{sec:discussion}

From the spectral analysis of the eight sources presented in this paper a diverse range of temporal characteristics have been identified in the bright (\LX $\ge1.2\times 10^{38}$\ergps) LMXB population of NGC 3379. Here the properties of the four sources determined to exhibit spectral variability are presented individually and their best-fit parameters are compared to the spectral simulations discussed in section \ref{sec:sims}. The spectral parameters of the remaining four sources are also presented and the overall properties of this bright population are discussed.

\subsection{A Black Hole GC-LMXB}

\begin{figure}
  \begin{minipage}{0.3\linewidth}
  \centering
  
    \includegraphics[width=\linewidth]{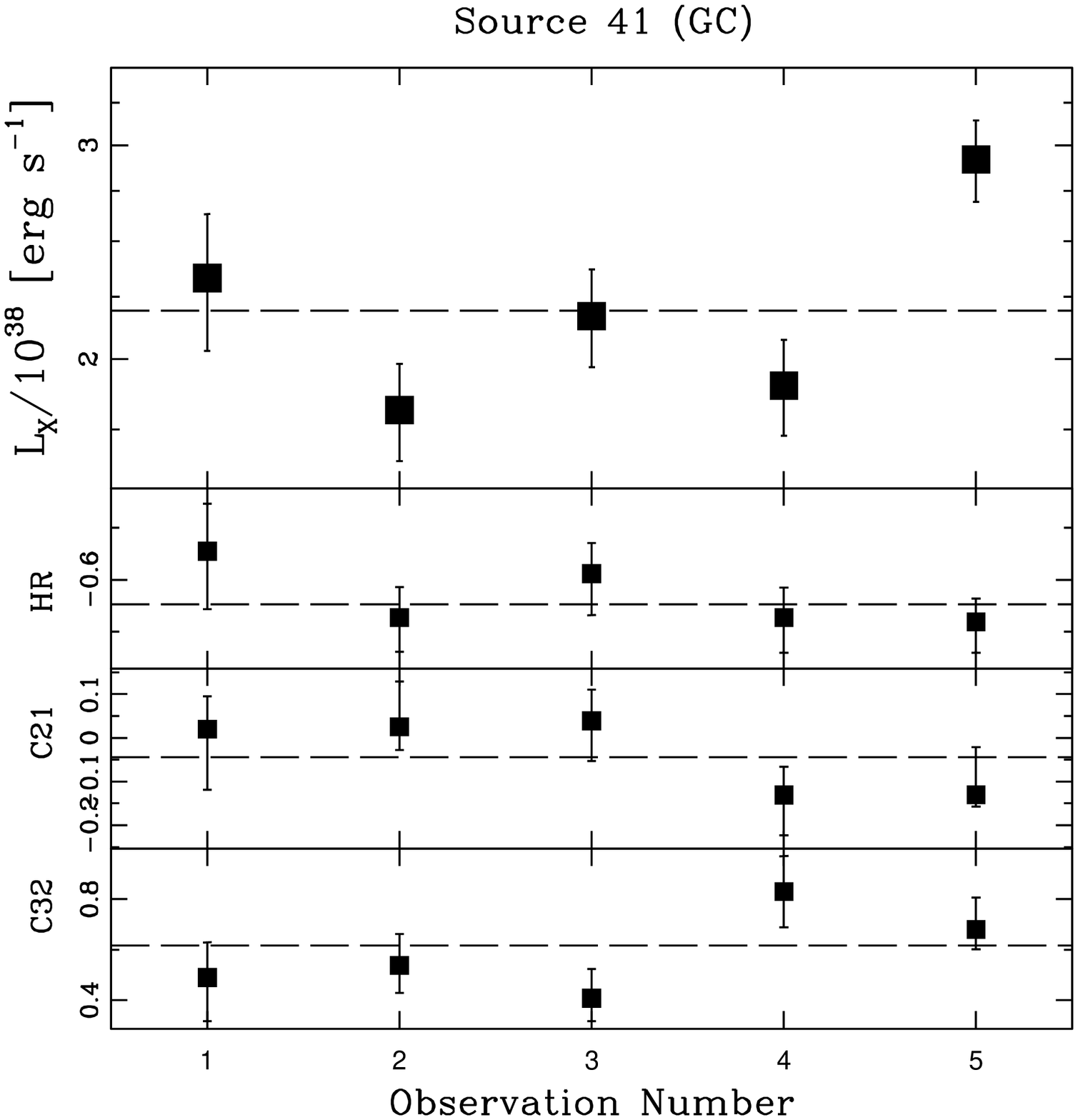}
  
  \end{minipage}\hspace{0.02\linewidth}
  \begin{minipage}{0.3\linewidth}
  \centering

    \includegraphics[width=\linewidth]{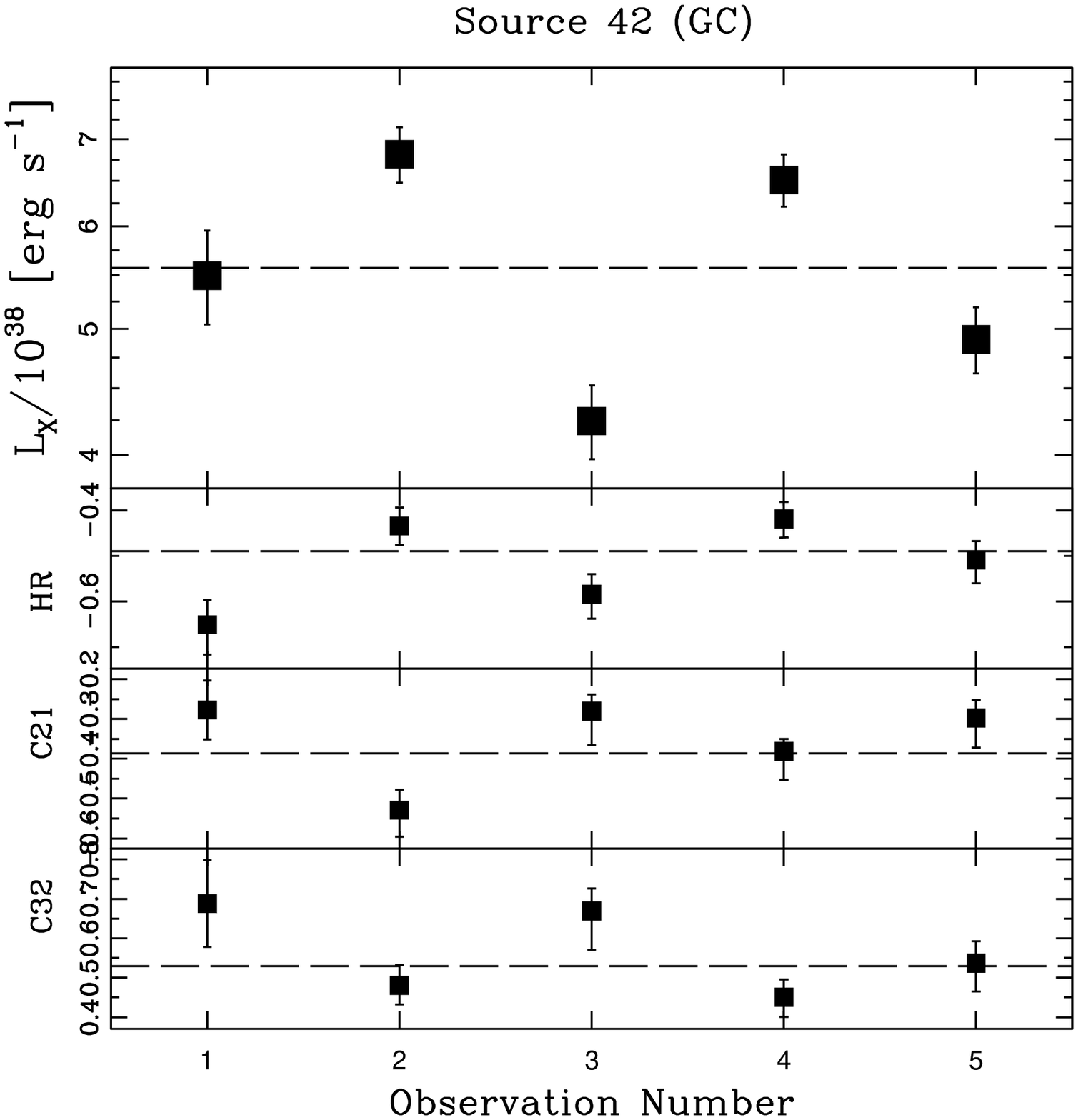}

\end{minipage}\hspace{0.02\linewidth}
\begin{minipage}{0.3\linewidth}
  \centering

    \includegraphics[width=\linewidth]{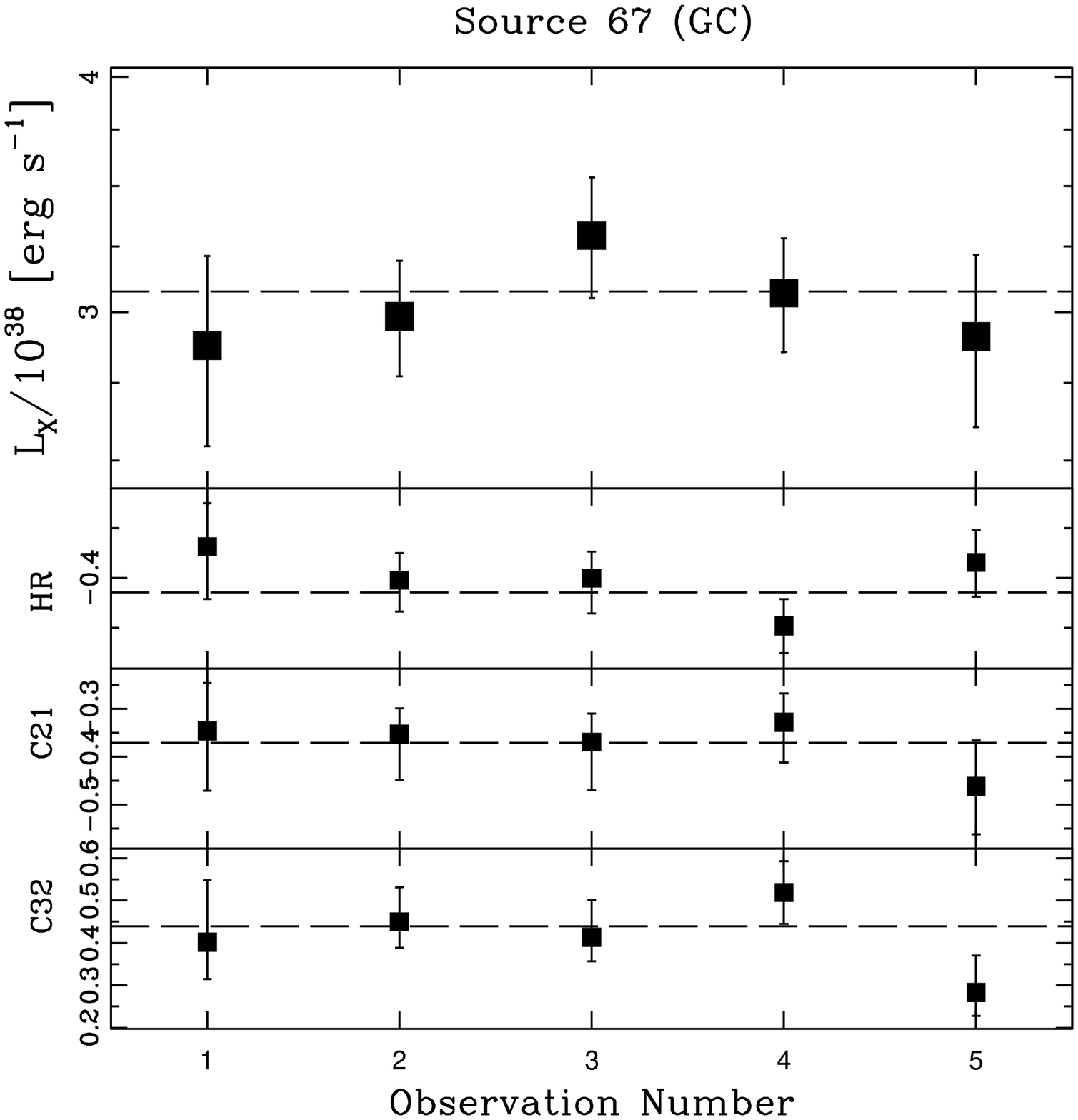}

 \end{minipage}\hspace{0.02\linewidth}

\begin{minipage}{0.3\linewidth}
  \centering
  
    \includegraphics[width=\linewidth]{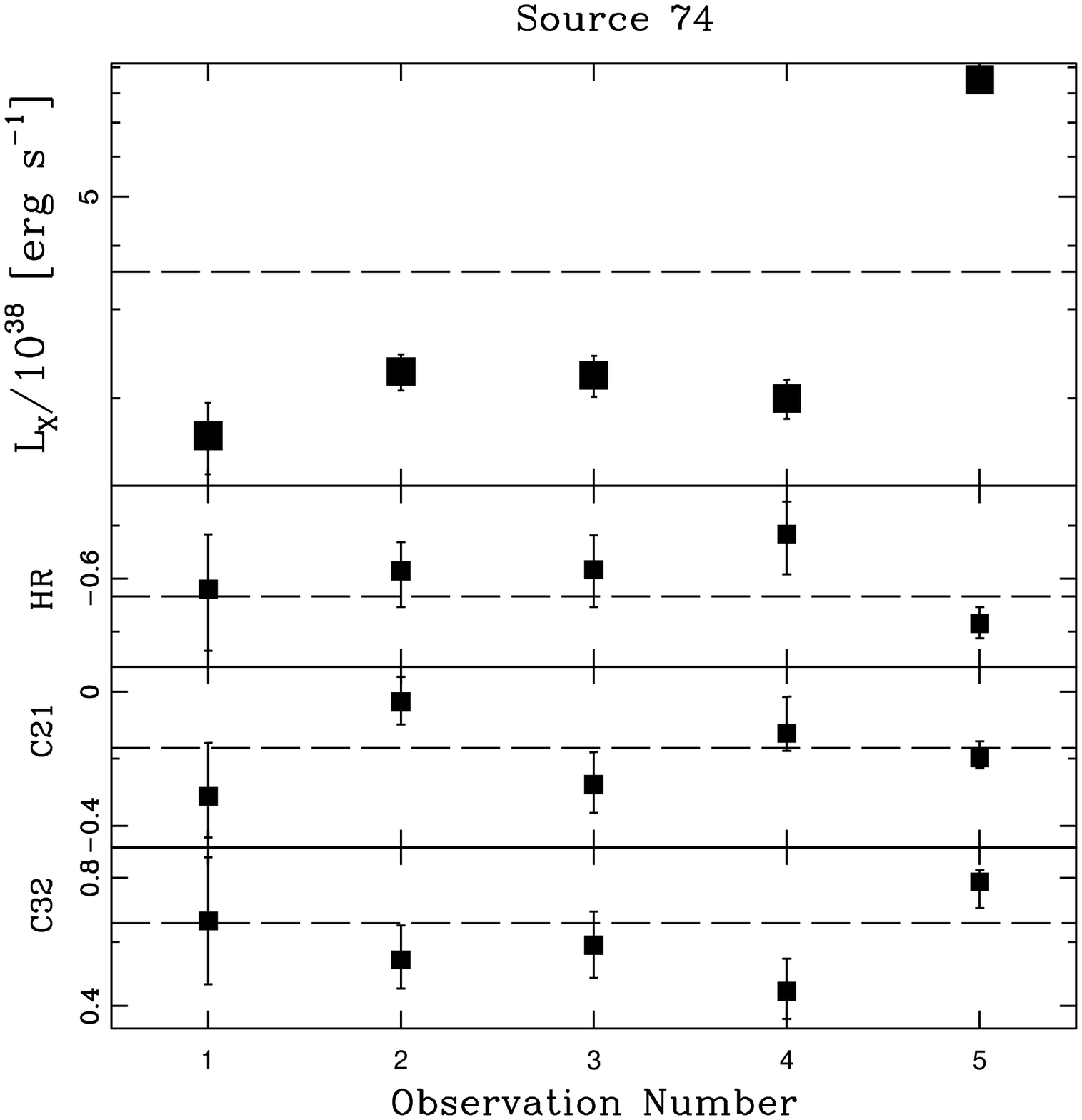}
  
  \end{minipage}\hspace{0.02\linewidth}
  \begin{minipage}{0.3\linewidth}
  \centering
  
    \includegraphics[width=\linewidth]{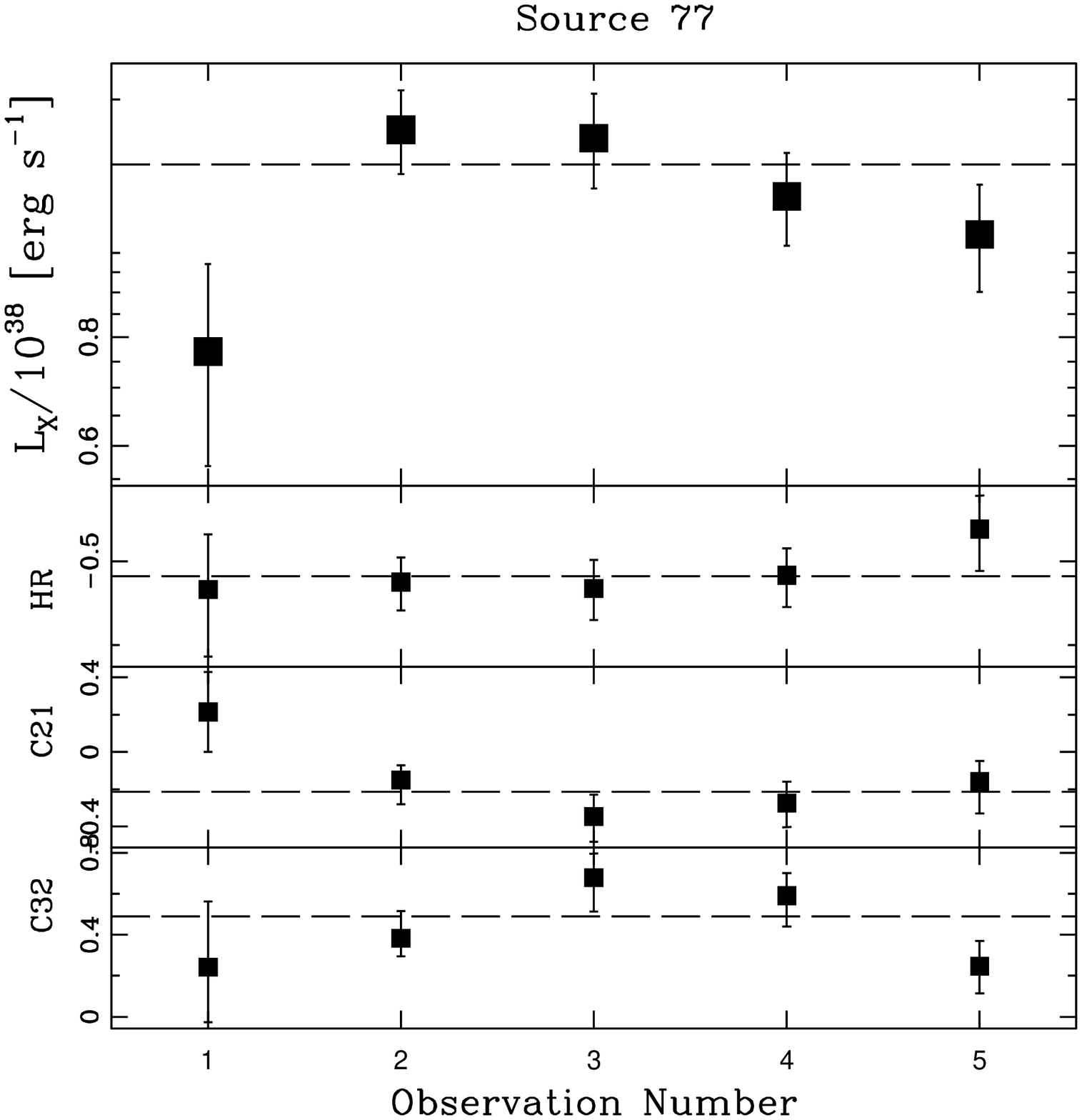}
  
  \end{minipage}\hspace{0.02\linewidth}
  \begin{minipage}{0.3\linewidth}
  \centering

    \includegraphics[width=\linewidth]{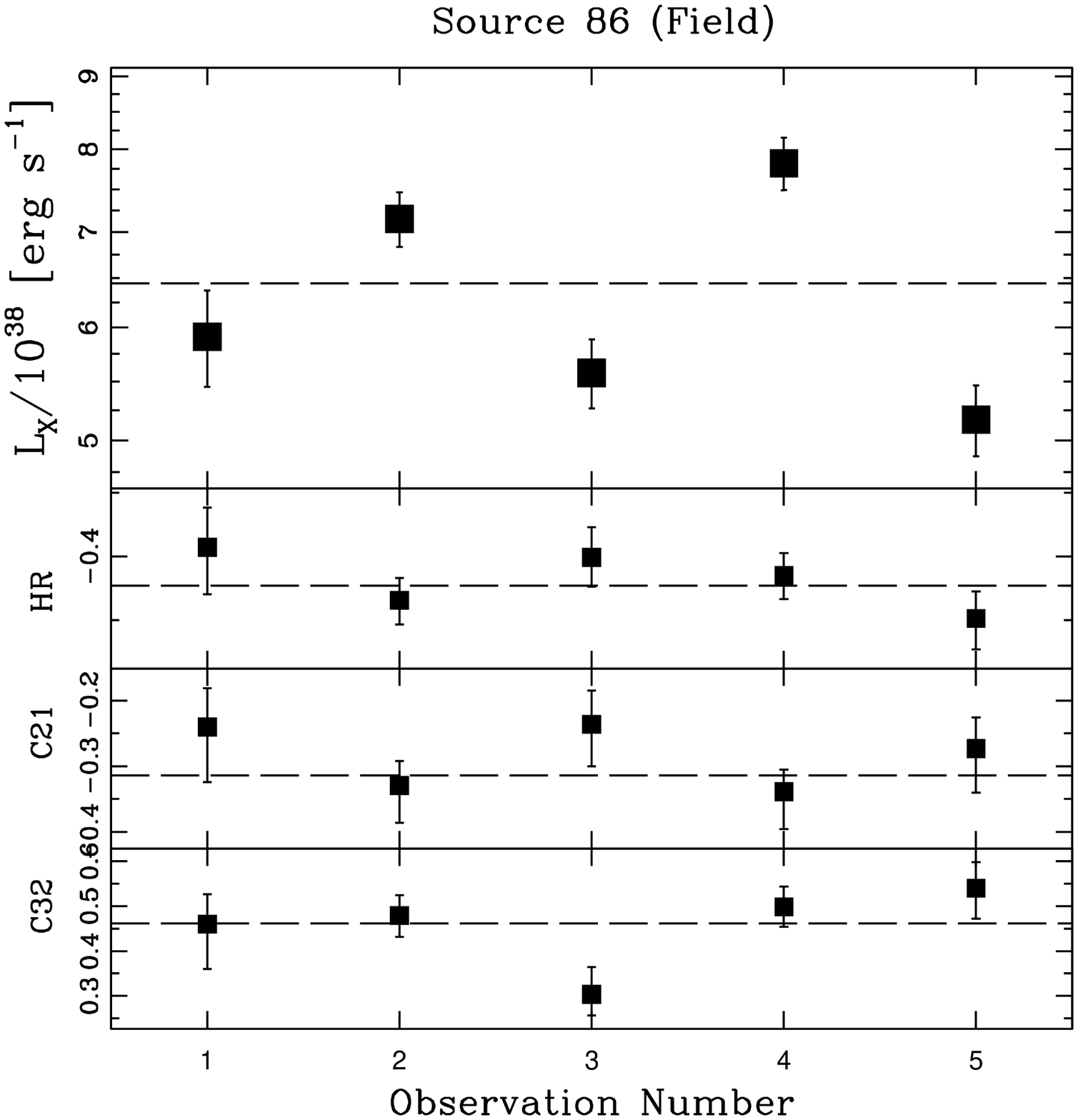}

\end{minipage}

\begin{minipage}{0.3\linewidth}
  \centering

    \includegraphics[width=\linewidth]{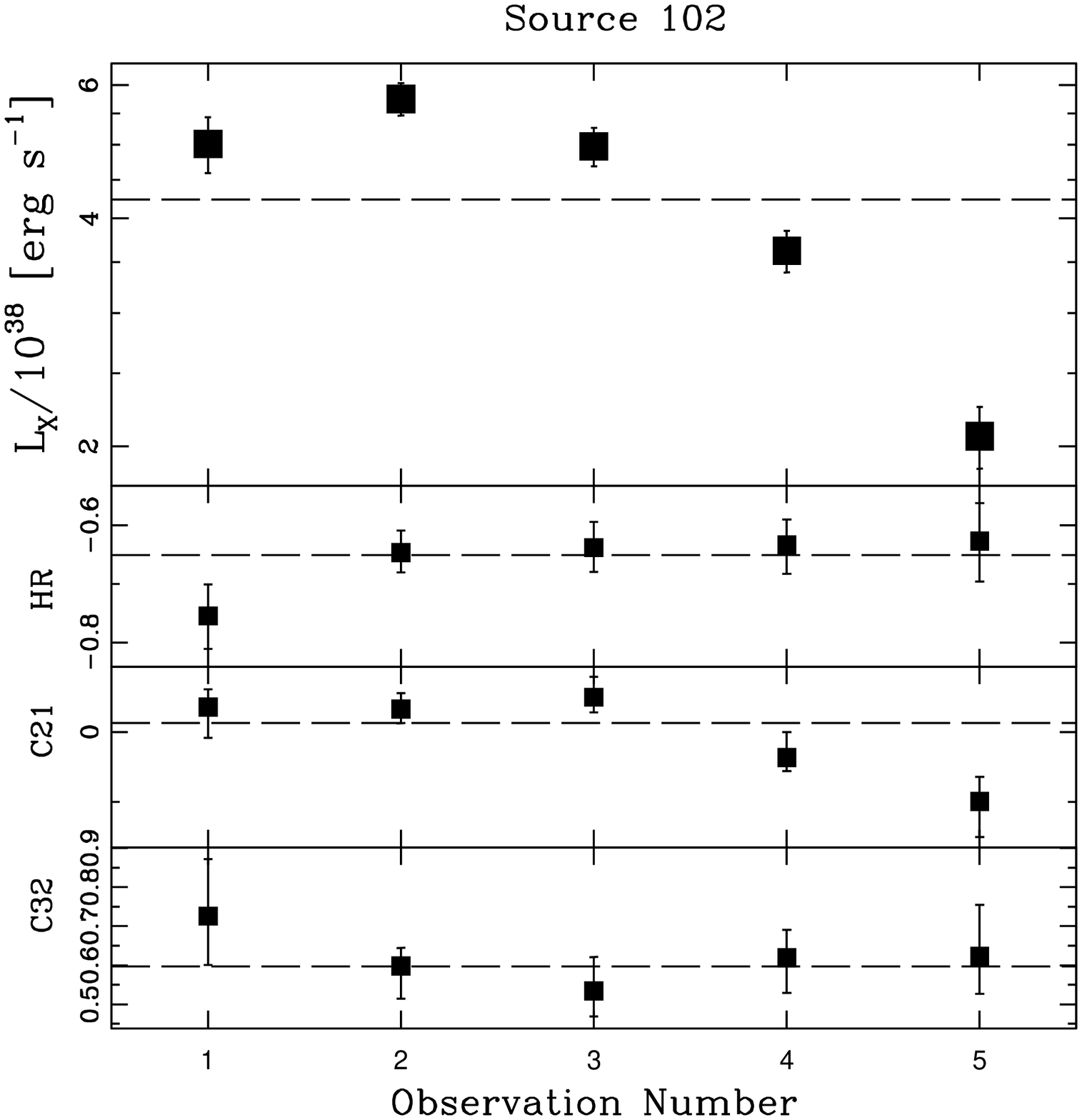}

 \end{minipage}\hspace{0.02\linewidth}
\begin{minipage}{0.3\linewidth}
  \centering
  
    \includegraphics[width=\linewidth]{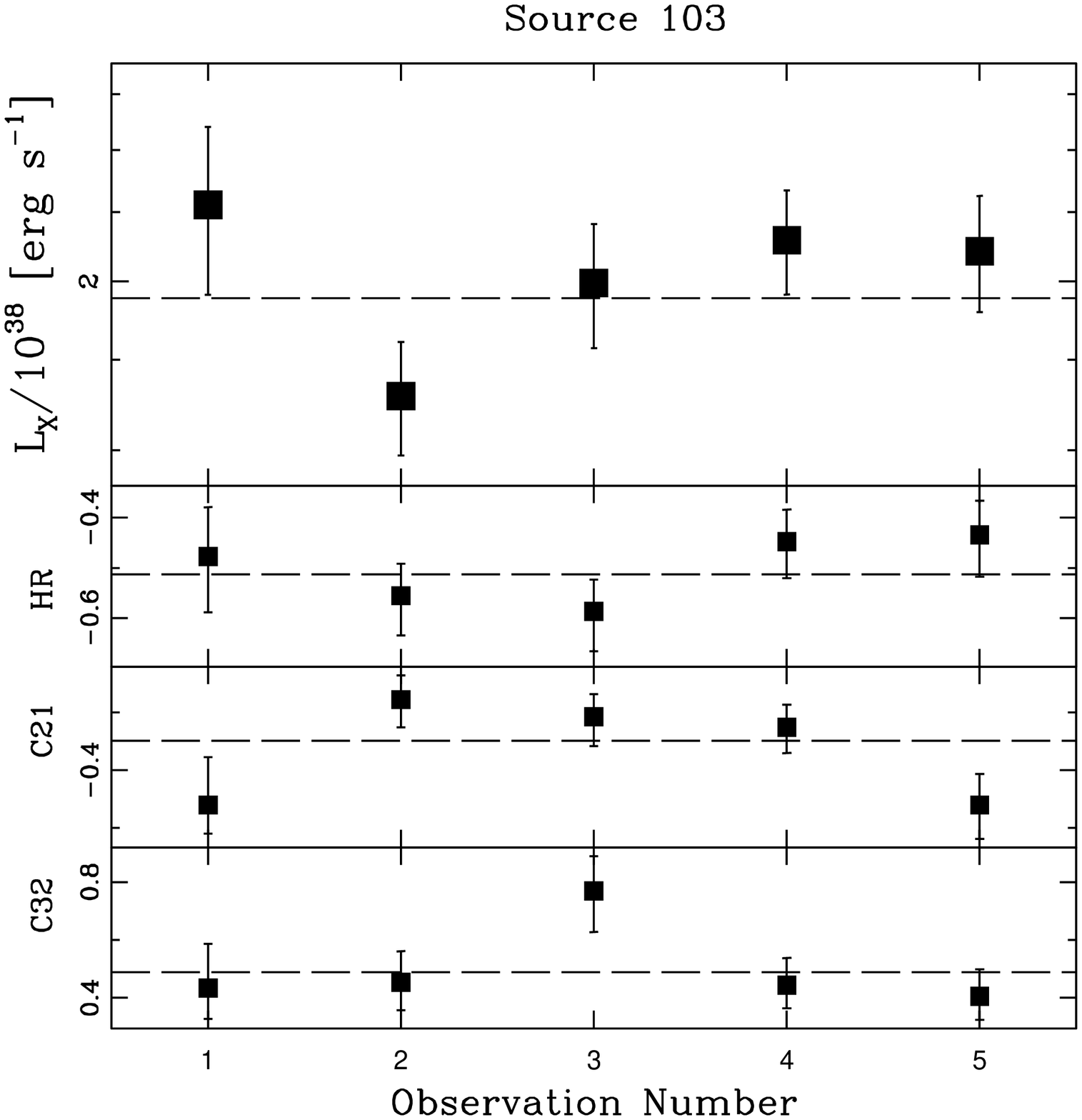}
  
  \end{minipage}\hspace{0.02\linewidth}
  \centering
  \caption{Plots of the eight bright sources in NGC 3379, adapted from B08. Each plot summarizes the
variations in properties of each source between each observation. In
the main panel the long-term light curves are shown. In the second
panel down, the hardness ratios are indicated. These are defined to be
HR = H$-$S / H+S, where H is the number of counts in the
hard band (2.0$-$8.0 keV) and S is the number of counts in the soft
band (0.5$-$2.0 keV). In the third and
fourth panels the color ratios; C21 and C32, are
plotted, where C21={\em log}S2+{\em log}S1 and C32=$-${\em log}H+{\em log}S2. For
the color ratios the bandwidths are defined to be S1=0.3$-$0.9
keV, S2=0.9$-$2.5 keV and H=2.5$-$8.0 keV. From B08 it was determined that all sources apart from S67 display long-term flux variability. \label{fig:lc} }

\end{figure}

Of the three GC-LMXBs that are presented in this paper, S42 is both the most luminous and the most variable. From Figure \ref{fig:lc} it can be seen that there is a significant increase (and subsequent decrease) in flux between observations implying $\Delta$\LX$\sim2.5\times10^{38}$ \ergps\ in the simple assumption of a canonical PO spectrum used to derive Figure \ref{fig:lc} (B08). This pattern of variability is fairly common for LMXBs (van der Klis 1994). The large flux variation observed very strongly indicates that S42 is not a superposition of multiple NS-LMXBs but a single source. In fact, if it is conservatively assumed that the total luminosity is from the contribution of multiple sources, the luminosity from a single source would have a lower limit of $\sim2.5\times10^{38}$ \ergps (increasing to $\sim7\times10^{38}$ \ergps\ if the emission arises from a single source). This luminosity is above the Eddington limit for a neutron star binary (although would be close to the limit of a heavy (2$-$3 \Msol; Kalogera \& Baym 1996) NS, or a 1.4 \Msol\ NS with a He or C/O donor) and indicates that the source is likely to be a black hole binary system.

Based on the X-ray luminosity, HR values and colors of this source, the spectra were separated into three different groups. In all three cases both single component models provided statistically acceptable fits to the data. In the PO model the spectral groups provided a range in $\Gamma$ of $\sim2.1-1.7$, which is within the range of a hard state BH-LMXB. However, in all three cases a large intrinsic \NH, of order 6 times the Galactic column, was required to provide an adequate description of the data. Such large \NH\ columns are not usually seen in LMXBs and would be especially difficult to explain in a sub-Eddington low-state source, where outflows are not expected. One explanation could be that the lower luminosity state is a consequence of foreground absorption from a warped disc (as advocated for the variable BH-GC source in NGC 4472, which showed variability consistent with arising from changes in \NH; Shih \etal\ 2008). However, since there is little variability detected in \NH\ between spectral groups for S42, this mechanism is unlikely to be the cause of the change in \LX. Instead, from the simulations presented in section \ref{sec:sims}, there is an indication that when a source has a significant flux contribution from a thermal disc component, the single-component PO model will provide an elevated absorption column, caused by the soft emission arising from the disc. Further, the DBB single-component modeling from the simulations indicates that when a source is in a TD state the best-fit absorption column $>$0, becoming consistent with Galactic absorption as the accretion disc completely dominates the flux contribution (see Figure \ref{fig:simsMCD}). 

Indeed, the DBB best-fit models of S42 suggests that this source is predominately in a TD state over observations 1$-$4, where \NH\ ranges between 0.6$-$3.2$\times 10^{20}$\pcmsq, possibly entering an intermediate state in observation 5, where its likely that the thermal and non-thermal flux contribute equally to the emission. This disc component in the TD state appears to be at its most dominant when the source is at its most luminous, in spectral grouping 2, where the best-fit value of \NH\ is consistent with Galactic and \kt$\sim$1.5 keV, which infers a mass of $\sim$4 \Msol\ (cf. Gierli\'nski \& Done 2004). 
This best-fit model provides a luminosity of $\sim9.1\times10^{38}$ \ergps, which further strengthens the interpretation that this source is a black-hole binary, where flux variability indicates that there must be a source with \LX$>4\times10^{38}$ \ergps\ within the GC. 
%
The increase in temperature coupled with the increase in luminosity observed between the spectral groupings could be explained as the disc's response to a temporary increase of the accretion rate, or could be due to the failure of the single-component model to adequately describe the properties of the spectra as the PO component becomes more prominent. However, the statistics in these spectra do not constrain more complex models (e.g. PO+DBB) to allow this to be investigated further. 

What is very likely from these spectral properties is that a single black-hole binary has been detected within a globular cluster in NGC 3379, in addition to the BH-GC within NGC 4472 (Maccarone \etal\ 2007) and the suggested IMBH-GC in NGC 1399 (Irwin \etal\ 2010). Further, it has also been determined that this source is in a thermally dominant state when it is at its most luminous.
Maccarone \etal\ (2007) find that the BH-GC in NGC 4472 is in a very luminous ($\sim7\times10^5$\Lsol) and metal poor GC and Irwin \etal (2010) determine that the source in NGC 1399 is in a luminous ($\sim3\times10^5$\Lsol) but metal rich GC. In NGC 3379 not only S42, but also the two other GC-LMXB sources; S41 and S67, are also found in luminous GCs (B08), with luminosities of order  1.5$\times10^5$\Lsol, 3$\times10^5$\Lsol\ and 2$\times10^5$\Lsol\ respectively, therefore following the trend identified in previous GC-LMXB studies (e.g. Kundu \etal\ 2007). Using the color-metallicity relations in Smits \etal\ (2006), and the GC colors reported in B08, we find that S41 and S67 are in `blue' low-metallicity clusters (V$-$I=0.85 and V$-$I=0.87; $\sim$1/10 solar), whereas S42 is in a `red' high metallicity cluster (V$-$I=1.12). 

These comparisons, albeit based on only five objects, are in agreement with the conclusion that GC mass may be a factor in retaining a BH-LMXB. However, given that this paper only concentrates on the brightest LMXBs within NGC 3379, we do not speculate on the suggestion that the more massive GCs host the more luminous LMXBs (Kundu \etal\ 2007). In fact, from B08 it has been determined that there are two less luminous LMXBs that reside in more luminous GCs than the sources presented here. Further, in the cases of S41, S67 and the source in NGC 1399, without large temporal luminosity variations it cannot be ruled out that these GCs may contain a superposition of less luminous NS rather than a single bright BH. However, due to the large variability in both S42 and the source in NGC 4472, it can be confirmed that these sources are bright single objects.

\subsection{A Field BH-LMXBs}

S86, the only confirmed field source with \LX$\ge10^{38}$\ergps\ within NGC 3379, exhibits similar \LX\ variability to that of the BH-GC, S42. However, in this case, when the source is observed in its less luminous state the spectra are consistent with either a hard state, where $\Gamma\sim$1.5 and no extra absorption is required, or possibly a cool-disc model. This latter interpretation is proposed as the best-fit value of \NH\ tends to 0 when left free to vary and the single-component DBB model is statistically rejected when applied to these spectra. When investigating the cool-disc simulations, such behavior was observed in both these single-component model fits (see section \ref{subsec:cool}). In the higher flux observations the DBB model describes the spectra well, where the best-fit inner disc temperature is $\sim1.4$ keV, consistent with the TD state of a $\sim$5 \Msol\ BH. The PO model also provides an acceptable description of the spectra in this higher flux state, with $\Gamma\sim$1.6, typical of hard state spectra. However, in this more luminous state, whilst the spectral index indicates a NT source, the absorption column is a factor of 4 greater than the Galactic \NH. This could again point to the single component model failing to describe the nature of the accretion, indicating a change in spectral state from the NT to the TD state (as investigated in the simulations in section \ref{sec:sims}), or it could indicate the presence of a warped disc, as discussed in Shih \etal\ (2008). In both models the source luminosity exceeds the NS Eddington luminosity, with \LX\ ranging between $\sim 7\times10^{38}$\ergps\ in the NT state (or cool-disc) observations (where this high \LX\ also indicates that this source is not residing in the NT state), to $\sim5\times10^{38}$\ergps\ in the TD state spectra, which is therefore consistent with emission from a stellar-mass BH.

\subsection{Spectral State Change of Central Region LMXB}

S74 resides in the central region of the galaxy where the optical data become confused and therefore has an unclassified optical correlation. The spectral behavior of this source is very stable in the first four observations, where the spectra are well fitted with a $\Gamma\sim$1.6 PO model with no intrinsic absorption, consistent with a NS or BH-LMXB in the hard-state. Although, this source could have a cool-disc component, given that \NH\ tends to 0 when left free to vary in the PO model, and the DBB model provides a poor description of the data. In observation 5 the spectral emission changes dramatically, increasing in flux and altering in spectral state, where the PO model indicates that the source has become softer and the absorption column has increased. This is again interpreted as an indication that the disc emission has become more prominent in the source, where the DBB model describes the spectra well, although \NH\ left free to vary provides a best-fit value well below Galactic and has therefore been frozen in the final model. These parameters indicate that the source is in a TD state, with an inner disc temperature of $\sim$0.7 keV and an intrinsic luminosity of $\sim 7.5\times10^{38}$\ergps, that, if following the $M\propto~T^{-4}$ relation, indicates that the source is a $\sim$75 \Msol\ black hole. However, as the absorption column from the DBB model is below the Galactic value it is possible that a non-thermal component is present in the source, therefore this derived mass should be treated as an upperlimit.

\subsection{A Luminous Cool-Disc Source}
\label{subsec:cooldisc}

The final source with spectral variability is S102, which is just outside the {\em HST} FOV. From the plots indicating the variability in \LX\ and HR (Figure \ref{fig:lc}) alone it appears that this source is gradually transitioning from a bright hard source to a lower luminosity soft source. However, from the spectral groupings it has been determined that the source actually resides in distinct states. In the first observation the spectrum is well described by a steep power law ($\Gamma\sim$2.6) with no intrinsic absorption. This photon index value is steeper than would be expected for a source in the hard-state (RM06: $\Gamma<$2.1), suggesting that the source is in a SPL state. However, given that the single-component DBB model is statistically rejected when fitted to this spectra, this source could be in an intermediate state with a cool-disc component (see section \ref{subsec:cool}). Although, within errors, both \NH\ and $\Gamma$ are consistent with the expected values of a SPL state source, and there are too few counts in the data to allow additional components to meaningfully constrain the spectral parameters.

In observations 2 \& 3 the spectra exhibit very similar properties to those observed in observation 1 although, with a greater number of counts, it is clear that the fit value of \NH\ from the PO model is significantly below that of Galactic absorption (see Figure \ref{fig:Src102con}) and consequently a two-component model has been applied to the spectra. From this best-fit model the thermal emission has been determined to have an inner-disc temperature of \kt$\sim$0.14 keV, with a photon index of 1.6. A large absorption column is required for this fit, although this is not well constrained and is consistent with Galactic absorption (see contour plots in Figure \ref{fig:Src1022con}). Such a spectrum is consistent with the properties observed in some ULXs (e.g Soria \etal\ 2007) and indeed, the intrinsic luminosity of S102 in observations 2 \& 3 is $\sim$1.1$\times10^{39}$\ergps, which does fall under the classification of a ULX, albeit at the very low end of the luminosity range. Further, the flux ratio from both of these observations indicates that 40\% of the emission arises from the disc-component, slightly higher than the contribution determined in Soria \etal\ (2007), where the thermal emission contributed 25\% of the observed flux. From the $M\propto$$T^{-4}$ relation the temperature of this fit implies an intermediate mass black hole of $\sim 5\times10^4$\Msol, although, as discussed in section \ref{subsec:cool}, this disc may appears cooler due to the prescence of a Comptonizing corona altering the properties of the disc (Done \& Kubota 2006) and the true mass of the black hole may be much lower.

In the final two observations of this source the luminosity decreases significantly and the flux is well described by a single-component PO model, with \NH\ consistent with Galactic absorption and $\Gamma\sim$1.85, indicating that the source is likely to be emitting in a NT state. This source therefore provides observational information of a black-hole binary source transitioning from a steep power law cool-disc state to a hard state, where the luminosity falls from $\sim$1.1$\times10^{39}$\ergps to $\sim3.8\times10^{38}$\ergps.


%
%

\subsection{The Overall Population}

In addition to the four sources that exhibit spectral variability, the four other sources included in this paper have also been analysed. From their spectral parameters, two are consistent with being in an NT state (S41 \& S77) with photon index value of $\sim$1.9 and $\sim$2.0. The two other sources (S67 \& S103) have large values of intrinsic \NH\ from the PO model, indicating that a significant disc is present in the spectra. In the case of S67, the DBB model provides a value of \NH\ consistent with Galactic absorption, suggesting that the source is in a TD state, whilst S103 has a value tending to 0 and is therefore likely to be in an intermediate state with flux contributions from both the thermal and non-thermal components of the binary. 

\begin{figure}
  \centering
\begin{minipage}{0.485\linewidth}
  \centering

    \includegraphics[width=\linewidth]{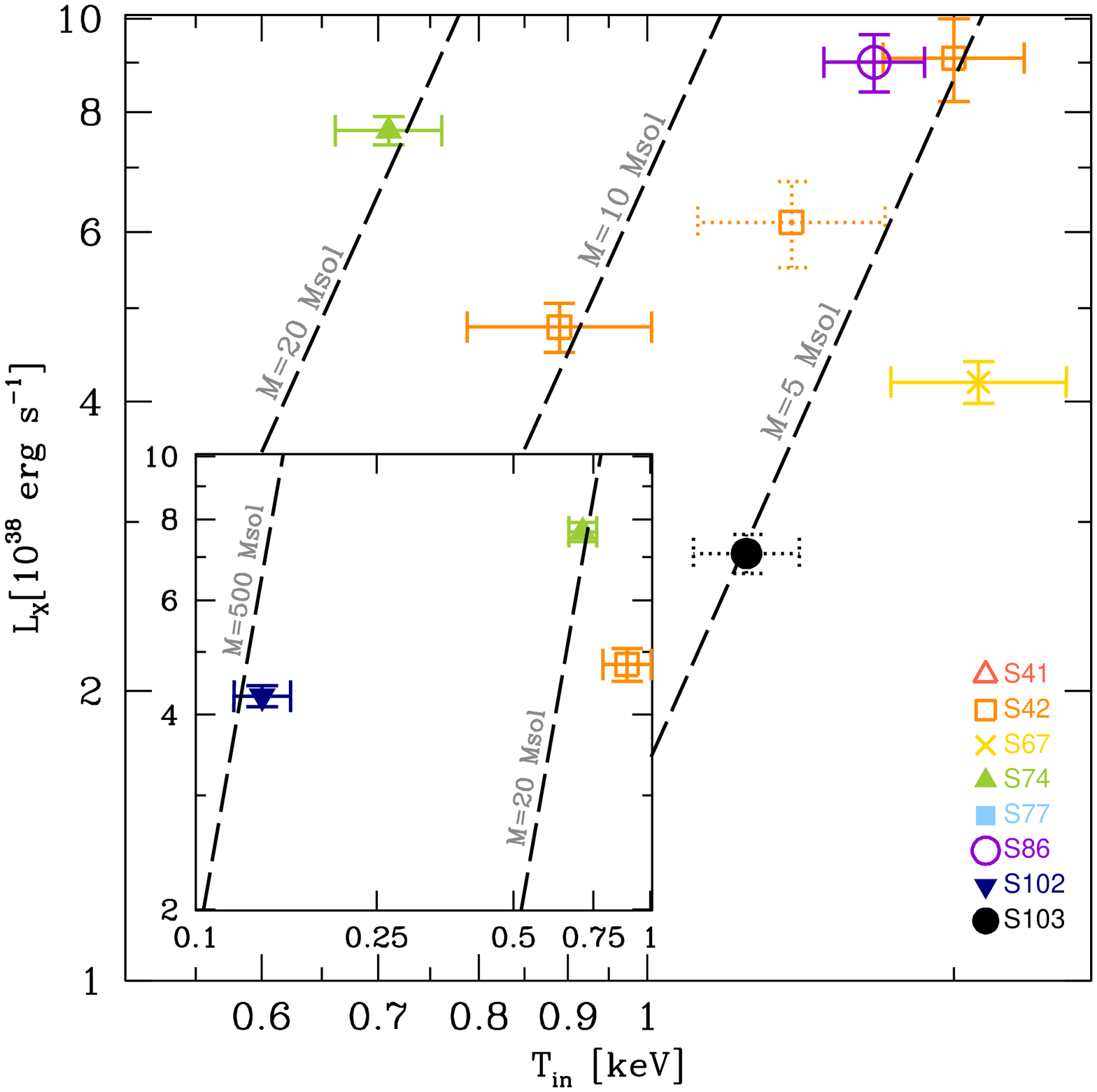}

 \end{minipage}\hspace{0.02\linewidth}
\begin{minipage}{0.485\linewidth}
  \centering

    \includegraphics[width=\linewidth]{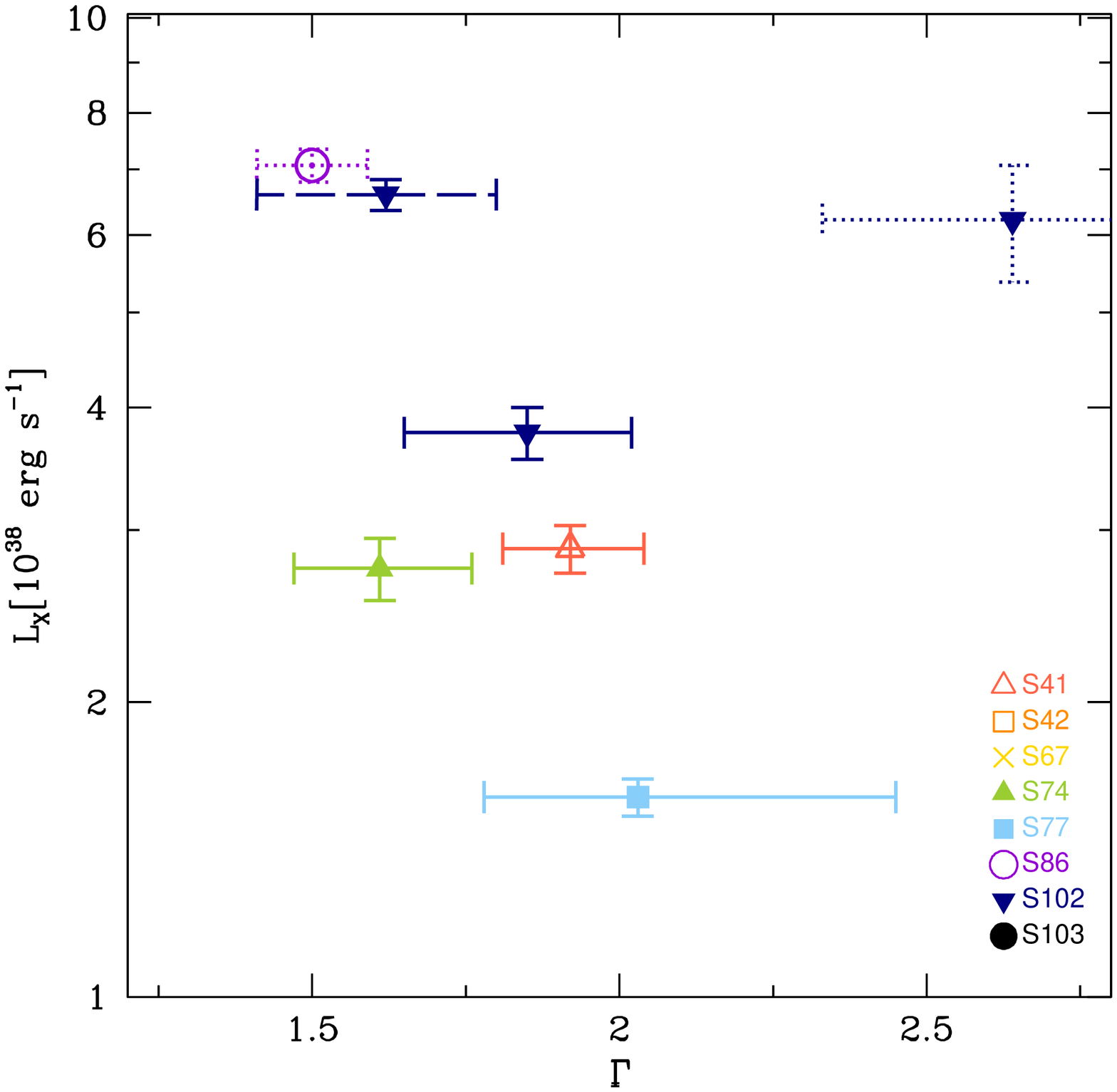}
 \end{minipage}\hspace{0.02\linewidth}

    \caption{A population plot presenting the \LX\ and \kt\ or $\Gamma$ values of the eight sources presented in this paper, where multiple points are plotted for spectrally variable sources to indicate the best-fit values from each spectral grouping. In this figure the left panel presents \LX$-$\kt\ for sources in a thermally dominant state and the right panel presents \LX$-\Gamma$ for sources in a hard state. In both panels the points represent the results from the spectral fits, with 1$\sigma$ errors indicated. Each source is denoted by symbol and color, with labeling provided in the bottom right corner of each panel. In cases where a source has been determined to be in an intermediate state dotted error bars are presented. In the case of observations 2\&3 for S102 the thermal and non-thermal components are plotted separately in both panels and are indicated by dashed error bars. In the \LX$-$\kt\ panel this point is not included in the main figure but within an insert, where, due to the cool temperature of the disc (0.14 keV), the x-axis has been adjusted to cover \kt\ values between 0.1$-$1.0 keV. In the main $L-$\kt\ plot the diagonal dashed lines indicate the $L\propto$$T^4$ relation for BH masses of 5, 10, and 20 \Msol, following the relation of Gierli\'nski \& Done (2004). In the insert plot, in addition to these relations a line indicating $L\propto$$T^4$ for a BH mass of 500 \Msol\ is also presented. \label{fig:pop} }

\end{figure}

In Figure \ref{fig:pop} the spectral properties of the binaries investigated in this paper are presented. In this figure the spectral groupings of each source have either been plotted in the left or right panel, depending on their inferred state, where in the left panel log \kt\ against log \LX\ is presented and the right panel shows $\Gamma$ against log \LX. When sources are likely to be in a spectral state that is not TD (or NT) the errors bars are presented as dotted lines. The best-fit values from the PO and DBB components of the luminous, cool-disc observations of S102 are indicated by the dashed error bars. In the case of the TD state for this source, this point is presented as an insert within the main plot, where a lower disc temperature range is presented.
In addition to the sources, the $L-T$ relations for a non-rotating BH (cf. Gierli\'nski \& Done 2004; eq. 3, assuming an inclination angle of 60 degrees and $f_{\rm{col}}$=1.8) of masses 5\Msol, 10\Msol\ and 20\Msol\ are indicated by the dashed lines in the left-hand plot, and an additional line indicating the  $L-T$ relation for a BH mass of 500\Msol\ is included in the insert plot. In both panels, sources are indicated by symbol and color where labels are given in the bottom right corner.

From comparing both panels it can be seen that, in general, when a source is emitting at a lower luminosity, the spectrum indicates that the source is in a NT state, dominated by the PO component. This behavior is commonly seen in Galactic binaries, where this state was previously defined as the low-hard state, denoting both the photon index and flux of the source. However, in McClintock \& Remillard (2006) it was advocated that luminosity should not be part of the classification scheme, as, while many Galactic binaries do become harder as their luminosity decreases, there are counter examples indicating that this behaviour is not ubiquitous (e.g. GX 339-4; RM06). Indeed, even within the sample of sources presented here there is an indication that we observe a source transitioning between a TD and NT state whilst exhibiting very little change in luminosity (S86), although it is possible that these NT states (obs 1, 3 \& 5) are actually in a cool-disc state. The other bright sources in the $\LX-\Gamma$ plot are observations 1 and observations 2\&3 of S102. From spectral modeling it has been determined that this source is actually in a steep power law or intermediate cool-disc state in observation 1 and in an intermediate cool-disc state in observations 2\&3, both of which typically result in a higher level of flux than the NT state.

In the left-hand panel the five sources determined to be in a TD or intermediate/TD state are presented (along with an insert presenting the TD properties of S102 in a luminous cool-disc state). The five sources within the main plot all reside in the same $L-T$ parameter space as Galactic black-hole binaries (cf. Miller \etal\ 2004; Kajava \& Poutanen 2009), albeit at the brighter end of the relation. S102, on the other hand, exhibits spectral properties that do not follow the typical relations observed in Galactic binaries (e.g. Miller \etal\ 2003; Stobbart \etal\ 2006) and could suggest the presence of an IMBH. However, as discussed in \textsection\ref{subsec:cooldisc}, with such a cool-disc, while it is consistent with this interpretation, it is not proof of an IMBH.

In this figure S42 is the only LMXB that has been determined to exhibit multiple spectral properties whilst in the TD state and is denoted here by the open squares. By comparing these points with the dashed lines on this plot, which indicate the $L$$\propto$$T^4$ relation; the expected signature for optically-thick, geometrically thin accretion discs (Done, Gierli\'nski, Kubota 2007), it appears that this source does not follow the standard relation, but instead roughly follows $L$$\propto$$T^{1.25}$. A flatter relation of $L$$\propto$$T^{2}$ has been observed in other black hole binary systems at higher temperatures ($\sim$1 keV) (Kubota \& Makishima 2004) which has been explained as the transition of the disc from a standard accretion disc (Shakura \& Sunyaev 1973) to a slim disc solution (Abramowicz \etal\ 1988) as the accretion rate increases. However, Gierli\'nski \& Done (2004) argue that such a transition should only occur above \Ledd, which is greater than the luminosity of sources $\sim$1 keV in Kubota \& Makishima (2004), and they instead suggest that this flattening is associated with a change in the color temperature correction ($f_{\rm{col}}$). 

In the case of S42, the inferred mass of the BH from the $M-T$ relation is $\sim$4\Msol, therefore the observed luminosity of 9.1$\times10^{38}$\ergps\ would require super-Eddington accretion for a BH of this mass. This suggests that the flatter relation in Figure \ref{fig:pop} is a consequence of the source transitioning to a slim disc, although a change in the color temperature correction cannot be ruled out. However, this suggestion of a flatter relation should be treated with caution due to, not only the limited data points, one of which is from the source in an intermediate state (observation 5) and therefore is likely to have an incorrect \kt, but also due to the uncertainty in \LX\ from the single-component spectral modeling. 

 



\section{Conclusions}
\label{sec:conclusions}

In this paper we have investigated the spectral properties of eight bright (\LX$>1.2\times^{38}$ \ergps) sources within NGC 3379, selected from the B08 catalog. From deep multi-pointing \CHANDRA\ observations we have been able to determine the source properties at different epochs, and thereby characterize source spectral variability. A set of simulations used to infer the spectral states of these sources has also been presented. 

Due to the luminosity cut imposed on this sample, all of the eight sources are either NS emitting at, or above the Eddington luminosity, or BHBs, with source luminosities ranging between $1.6-10.8\times10^{38}$\ergps. From optical correlations three sources have been determined to be BH-GCs, one is detected in the field, while the remaining four sources have insufficient optical data to classify a correlation. In all cases (bar S102) single component PO and DBB models have been applied to the data and have provided statistically acceptable fits.

\begin{figure}
  \centering
    \includegraphics[width=0.95\linewidth]{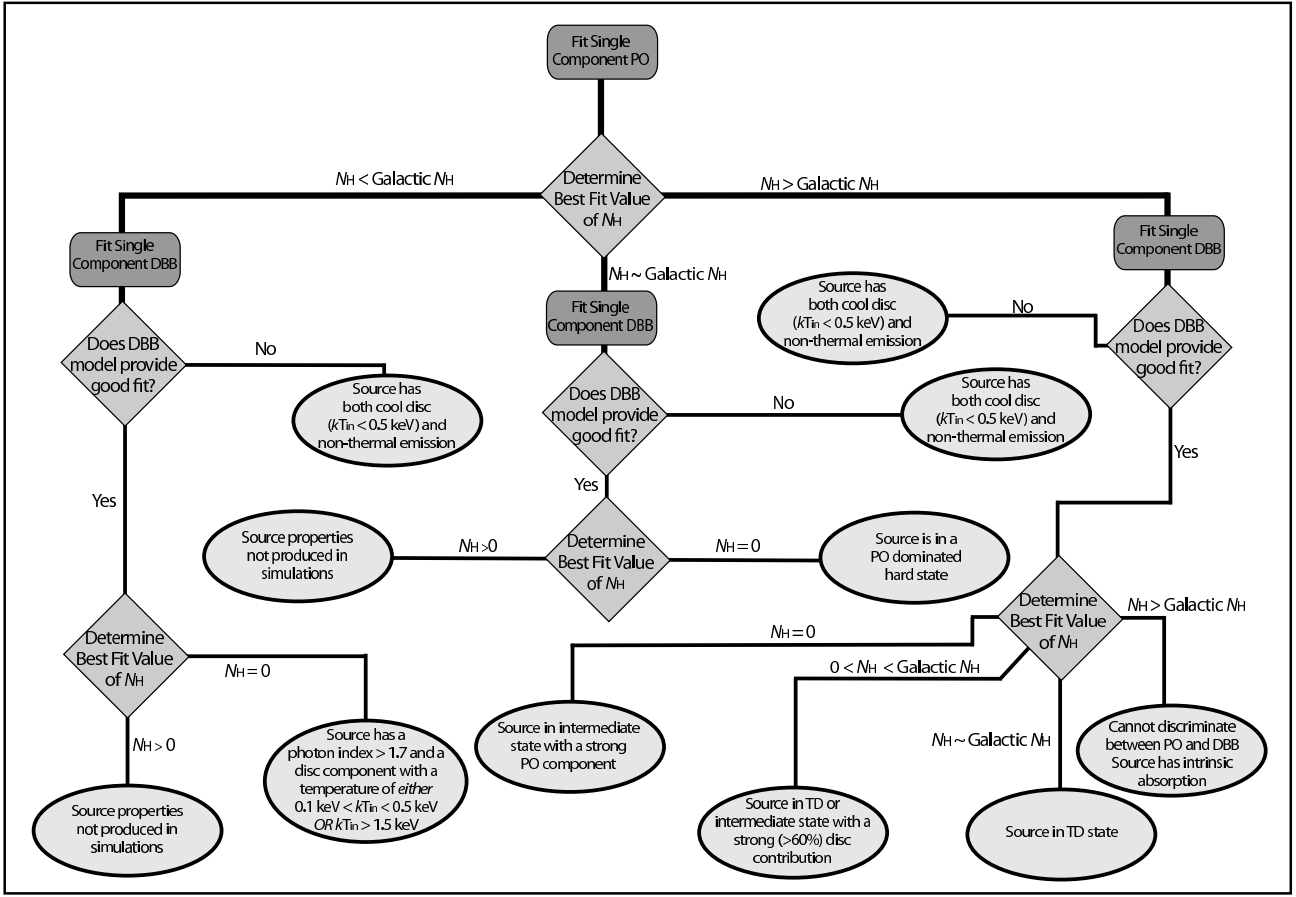}
    \caption{A decision tree summarizing how the best-fit absorption values from the PO and DBB single-component models from the simulations can be used to determine spectral source states. In this figure the interpretation of each combination of \NH$_{(PO)}$ and \NH$_{(DBB)}$ is presented in the ellipses, where TD, NT, and intermediate source states are given, including sources with cool-disc components.  \label{fig:flow} }

\end{figure}

To aid the interpretation of these single component models, simulations of low-count (250, 500 and 1000) data have been presented. This work demonstrates that it is possible to determine if a source is in a hard, thermally dominant, or intermediate state by comparing the best-fit values of \NH\ from the two spectral models. In cases where the line-of-sight absorption in the PO fit is significantly higher than that of the Galactic value it is likely that a strong thermal component is present. If it is then determined from the DBB model, that this same source has a best-fit \NH\ consistent with the Galactic value it points to the source being in a TD state. If instead the DBB best-fit \NH\ tends to 0, it is likely that the source has both strong thermal and non-thermal components and can be said to be in an intermediate state. Although we add the caveat that whilst an elevated value of \NH\ from the single-component PO model is indicative of a thermally dominant state, intrinsic absorption, arising from the disc could also explain this result in some cases. We have also provided simulations of cool-disc spectra, allowing us to identify when a source is in a high accretion state with a cool-disc component.
A summary of how the best-fit values from single-component models can be interpreted is provided in Figure \ref{fig:flow}.

From these simulations it has been determined that when a source has prominent disc emission as well as a significant non-thermal component, the luminosity derived from the single-component DBB model should be taken as a lower limit of \LX. The value inferred from the single-component PO model will be much greater than the true source value and should be disregarded. However, when a canonical PO model ($\Gamma$=1.7 and \NH\ frozen to the Galactic absorption value) is applied to the data, the derived \LX\ value will provide the closest value to the true source luminosity. When the source is in a TD state the X-ray luminosity from the DBB should be adopted. 

We have also performed these simulation for a range of different absorption values, confirming that the results are not significantly dependant on the value of Galactic absorption, demonstrating that these simulations can be used to interpret a variety of extra-galactic binary spectra. Further, different absorption models have also been been used in a sub-set of simulations, and these results have confirmed that with low-count data the choice of absorption model does not affect the interpretation of spectral state.

Four of the eight sources presented in this paper have been determined to exhibit spectral variability and observations have been subsequently been grouped to reflect these variations. By comparing the spectral properties of these groupings to our simulations, spectral transitions have been identified. The most luminous source detected within a GC, S42, has been determined to show significant variability (\LX$>4\times10^{38}$\ergps) and can therefore be confirmed as a strong BHB candidate, predominantly residing in the thermally dominant state. The best-fit inner-disc temperature of this source is 1.5 keV, and we estimate a black hole mass of $\sim$4\Msol, providing further evidence (in addition to Maccarone \etal\ 2007 and Irwin \etal\ 2010) that GCs do retain black hole binaries. If we assume that the two further GC-LMXB sources in this paper are also individual black holes, this indicates that 30\% of the GC-LMXB correlations in NGC 3379 contain a black hole binary, which constitutes a significant fraction of the GC-LMXB population.

A transition from a luminous steep power law state to a hard state has been identified in S102 where we have identified a binary with a cool-disc component (\kt=0.14 keV and $\Gamma$=1.6) emitting around the Eddington luminosity of a 10\Msol\ black hole (\LX$\sim 1\times10^{39}$\ergps). Taken at face value, the temperature of this cool-disc could suggest an IMBH. This source then declines over the three months between pointings to a source in a hard state, with $\Gamma \sim1.85$ and \LX=3.8$\times10^{38}$\ergps, consistent with NT state Galactic black hole binaries (RM06). The bright state of this source is similar to spectra observed in some ULXs (Soria \etal\ 2007), and indicates the similarities between these bright sources and more `normal' stellar mass black hole binaries in high accretion states.

The spectral properties of the eight sources are largely consistent with the parameters that have been observed in Galactic LMXBs, with sources in a NT state exhibiting a range of $\Gamma$=1.5$-$1.9 and sources in a TD state with \kt=0.7$-$1.55 keV. We have also shown that this population is consistent with the general trend of increasing luminosity as sources become softer, transitioning from a NT to TD state. 
The $L-T$ relation has been investigated for all the sources in a TD state, with the parameter space of these LMXBs (excepting the luminous cool-disc state of S102) being consistent with Galactic observations, albeit at the bright end of the $L\propto T^4$ relation. S42, the BH-GC, was the only source to be observed in the TD state in multiple pointings. The $L-T$ relation of this object is flatter than the typical relation, with luminosity increasing with temperature as $\sim~T^{1.25}$. Flattening has been observed in Galactic black hole binaries (Kubota \& Makishima 2004) and it has been suggested that this could be due to the standard Shakura-Sunyaev accretion disc evolving into a slim disc, although this flattening could also be explained by a change in the disc color temperature relation.


\acknowledgments

We thank the CXC DS and SDS teams for their efforts in reducing the data and 
developing the software used for the reduction (SDP) and analysis
(CIAO). We would also like to thank the anonymous referee whose detailed and careful
report has helped to improve this paper.
This paper is based upon work performed by S. Blake while visiting CfA as part of a student program sponsored by the University of Southampton.
This work was supported by {\em Chandra} G0 grant G06-7079A
(PI:Fabbiano) and subcontract G06-7079B (PI:Kalogera). We acknowledge
partial support from NASA contract NAS8-39073(CXC). A. Zezas
acknowledges support from NASA LTSA grant NAG5-13056.

{}

\clearpage

\LongTables

\begin{deluxetable}{cccccccc}
\tabletypesize{\scriptsize}
\tablecolumns{8}
\tablewidth{0pt}
\tablecaption{Observation Log and Source Counts \label{tab:log}}
\tablehead{\colhead{} & \colhead{Obs 1} & \colhead{Obs 2} & 
\colhead{Obs 3} & \colhead{Obs 4} & \colhead{Obs 5} & \colhead{All} & \colhead{Opt. corr} }
\startdata
OBSID		&	1587		&	7073		&	7074		&	7075		&	7076		&	-	&	\\
Exposure (ksec)	&	29.0		&	80.3		&	66.7		&	79.6		&	68.7		&	324.2	&	\\
\hline
\colhead{Net Counts} & \colhead{} & \colhead{} & 
\colhead{} & \colhead{} & \colhead{} & \colhead{} & \colhead{}\\
\hline 
S41	&	7.5$\pm$10.0  	&	146.0$\pm$13.5  &	143.6$\pm$13.2  &	150.0$\pm$13.6  &	200.0$\pm$15.4  &	718.7 $\pm$28.4	& GC	\\
S42 	&	170.4$\pm$14.1 	&	457.8$\pm$22.5	&	269.0$\pm$17.6 	&	 512.8$\pm$23.8 &	331.5$\pm$19.4 	&	1741.8$\pm$43.0 & GC	\\
S67	&	95.4$\pm$11.1  	&	239.5$\pm$16.9  &	219.0$\pm$16.1  &	243.2$\pm$16.9  &	198.7$\pm$20.9  &	1008.9$\pm$33.4 & GC	\\
S74	&	55.8$\pm$9.0  	&	181.1$\pm$14.8 	&	147.5$\pm$13.7  &	158.2$\pm$14.0  &	580.7$\pm$25.5  &	1165.4$\pm$35.9 & -	\\
S77	&	25.5$\pm$6.7   	&	111.1$\pm$12.2  &	90.0$\pm$11.2  	&	92.0$\pm$11.3  	&	71.7$\pm$10.1  	&	414.4$\pm$22.5  & -	\\
S86 	&	195.7$\pm$15.2 	&	572.9$\pm$25.3  &	371.4$\pm$20.6  &	620.4$\pm$26.2  &	352.4$\pm$20.2  &	2114.3$\pm$47.7 & F	\\
S102 	&	165.5$\pm$14.0  &	459.4$\pm$22.6  &	332.5$\pm$19.5  &	288.3$\pm$18.2  &	140.2$\pm$13.2  &	1389.0$\pm$38.7 & -	\\
S103	&	75.1$\pm$10.6 	&	131.3$\pm$12.6  &	127.1$\pm$13.4 	&	170.7$\pm$15.1 	&	143.1$\pm$14.1 	&	636.9$\pm$26.6  & -	\\
\enddata
\tablecomments{Optical correlations: GC indicates an association with a globular cluster, F indicates a field source, which is within the {\em HST} FOV and has no optical association. - indicates that there is insufficient optical coverage for this source to determine an optical association. All of these classification are fully described in B08, \textsection 2.6. }
\end{deluxetable}

\begin{deluxetable}{cccccccc}
\tabletypesize{\scriptsize}
\tablecolumns{8}
\tablewidth{0pt}
\tablecaption{Summary of Long- and Short-term Source Variability \label{tab:var}}
\tablehead{ \colhead{Source No.} & \multicolumn{5}{c}{Short-term variability} &\multicolumn{2}{c}{Long-term variability}  \\
\colhead{} & \colhead{Obs 1} & \colhead{Obs 2} & 
\colhead{Obs 3} & \colhead{Obs 4} & \colhead{Obs 5} & \colhead{Var.} & \colhead{Sign.} }
\startdata
S41		 		&	-	  	&	-		&	-		&	-		&	-		&	V	&	3.9$\sigma$	\\
S42 	 	 		&	-	  	&	-		&	-		&	-		&	-		&	V	&	5.9$\sigma$	\\
S67	 	 		&	-	  	&	-		&	-		&	-		&	-		&	N &		1.0$\sigma$	\\
S74	 	 		&	P	  	&	-		&	-		&	-		&	-		&	V &		15.8$\sigma$	\\
S77	 	 		&	-	  	&	-		&	-		&	-		&	-		&	V &		2.4$\sigma$		\\
S86 	 	 		&	-	  	&	P		&	-		&	-		&	-		&	V &		6.0$\sigma$	\\
S102 	 	 		&	-	  	&	-		&	-		&	-		&	-		&	V	&	10.7$\sigma$	\\
S103	 	 		&	-	  	&	-		&	P		&	-		&	-		&	V	&	2.0$\sigma$	\\
\enddata
\tablecomments{Individual variability is based on the Kolmogorov-Smirnov test: - indicates no variability, V, indicates variable source (with values $>$99\% confidence) and P indicates possible variable sources (with variability values $>$90\% confidence). The long-term variability from all five observations was defined by the chi-squared test where V indicates variability, and N indicates a non-variable source. Sign indicates the significance in change in luminosity between the highest and lowest flux observations. All of these classification are fully described in B08, \textsection 2.4. }
\end{deluxetable}

\clearpage

\begin{deluxetable}{c@{}c@{}c@{}ccccccccc}
\tabletypesize{\scriptsize}
\tablecolumns{12}
\tablewidth{0pt}
\tablecaption{ Summary of the best-fit parameters of the source spectra. Errors are given to 1$\sigma$ \label{tab:Bestfit}}
\tablehead{ \colhead{Source} & \colhead{Obs. No.} & \colhead{Net Counts} & \colhead{Model} & \colhead{$\chi^2/\nu$} &
\colhead{$P_{\chi^2}$} & \multicolumn{3}{c}{Parameters} & \colhead{\ensuremath{L_{\mathrm{X}}}} & \colhead{\ensuremath{L_{\mathrm{X}}} Range}  \\
\colhead{}  & \colhead{} & \colhead{} & \colhead{} & \colhead{}  & \colhead{} &
\colhead{\ensuremath{N_{\mathrm{H}}}} & \colhead{$\Gamma$}  & \colhead{\kt} &
\colhead{$\times 10^{38}$ erg $s^{-1}$} & \colhead{}  \\
\colhead{}  & \colhead{}  & \colhead{}  &
\colhead{}  & \colhead{}   & \colhead{} &
\colhead{$\times 10^{20}$} & \colhead{}  & \colhead{} &
\colhead{(0.3$-$8.0 keV)} & \colhead{}}
\startdata
S41$^{GC}$	& All	& 719 & PO	& 25.89/24    & 0.36	& 2.8Fr 				& 1.92$^{+0.12}_{-0.11}$	& -	& 2.87 	& 1.99$-$4.51			\\
\\
S42$^{GC}$	& 1\&3	& 439 &  PO	& 10.32/15	& 0.79	& 17.8$^{+4.6}_{-5.0}$	& 2.09$^{+0.10}_{-0.16}$	& -	& 5.60	& 4.21$-$8.15			\\
	& 1\&3	& 439 &  DBB		        & 12.00/16	& 0.74	& 2.8Fr				& -				& 0.89$^{+0.16}_{-0.05}$	& 4.78	& 3.18$-$6.08 \\
	& 1\&3	& 439 &  DBB		        & 11.32/15	& 0.73	& 0.6$^{+2.1}_{-0.6}$	& -				& 0.97$^{+0.13}_{-0.10}$	& 4.15	& 3.58$-$6.55 \\
\\
	& 2\&4	& 971 & PO	        & 32.86/40	& 0.78	& 18.8$^{+2.6}_{-2.5}$	& 1.68$^{+0.07}_{-0.10}$	& -	& 12.30	& 12.22$-$12.38			\\
	& 2\&4	& 971 & DBB		        & 40.05/40	& 0.47	& 3.2$^{+2.1}_{-2.1}$	& -				& 1.50$^{+0.16}_{-0.13}$	& 9.10	& 7.79$-$10.76 \\
\\
	& 5	&  332 & PO	        & 3.85/12	& 0.98	& 12.1$^{+4.2}_{-7.5}$	& 1.75$^{+0.13}_{-0.13}$	&  - 				& 8.26  & 7.80$-$9.60 \\
	& 5	&  332 & DBB	        & 5.44/13	& 0.96	& 2.8Fr				& -				& 1.21$^{+0.12}_{-0.17}$	& 6.14 & 3.99$-$6.94   \\
\\

S67$^{GC}$	& All	& 1002 & PO	& 31.50/34	& 0.59	& 17.4$^{+7.3}_{-6.0}$	& 1.57$^{+0.17}_{-0.14}$	& -				& 4.48 & 2.96$-$5.65   \\
	& All	& 1002 & DBB		        & 29.92/35	& 0.71	& 2.8Fr				& -				& 1.55$^{+0.11}_{-0.20}$	& 4.19 & 2.71$-$5.16   \\
	& All	& 1002 & DBB		        & 29.83/34	& 0.67	& 1.8$^{+4.0}_{-1.8}$	& -				& 1.55$^{+0.18}_{-0.17}$	& 4.67 & 2.82$-$5.09   \\
\\
S74	&123\&4	& 553 & PO	        & 16.22/16	& 0.45	& 2.8Fr				& 1.61$^{+0.15}_{-0.14}$	& -				& 2.74 & 1.51$-$3.55 \\
	&123\&4	& 553 & DBB		        & 26.39/16	& 0.05	& 2.8Fr				& -				& 1.01$^{+0.12}_{-0.18}$	& 1.95 & 0.97$-$2.38  \\
\\
	& 5  & 555 & PO	                & 35.53/21	& 0.03  & 11.6$^{+3.5}_{-4.0}$	& 2.13$^{+0.11}_{-0.14}$	& -			& 12.39 & 11.59$-$13.74 \\
	& 5  & 555 & DBB	                & 35.68/22	& 0.03	& 2.8Fr				& - 			& 0.71$^{+0.05}_{-0.05}$ 	& 7.65 & 6.63$-$7.89   \\
\\
S77     & All   & 368 & PO               & 6.72/7     & 0.46  & 21.9$^{+23.9}_{-20.7}$        & 2.03$^{+0.25}_{-0.42}$	& -                             & 1.60 & 0.37$-$2.20 \\
	& 	&     & DBB		        & 7.61/8	& 0.47	& 2.8Fr              		& -				& 0.98$^{+0.21}_{-0.13}$        & 1.28 & 0.49$-$1.71  \\
\\
S86$^{F}$& 13\&5	& 920 & PO	& 35.50/37	& 0.54	& 2.8Fr				& 1.50$^{+0.09}_{-0.09}$	& -		 	& 7.07 & 6.15$-$7.95 \\
\\
	& 2\&4	&1193 & PO	        & 39.38/54	& 0.86	& 13.1$^{+2.4}_{-2.2}$	& 1.65$^{+0.10}_{-0.08}$	& -				& 11.82 & 10.00$-$13.41  \\
	& 2\&4	&1193 & DBB		        & 42.17/51	& 0.81	& 2.8Fr				& -				& 1.35$^{+0.07}_{-0.11}$	& 9.01 & 6.73$-$10.63  \\
\\
S102	& 1	& 166 & PO	        & 194C  & 67\%C & 3.6$^{+4.7}_{-3.5}$		& 2.64$^{+0.39}_{-0.31}$	& -				& 6.22 & 5.30$-$7.07  \\
\\
	& 2\&3	& 792 & PO	        & 36.86/32	& 0.25	& 2.8Fr			& 2.11$^{+0.09}_{-0.09}$	& -				& 6.13 & 5.09$-$7.03  \\
	& 2\&3	& 792 & PO+DBB	        & 20.66/28	& 0.74	& 11.3$^{+16.0}_{-11.0}$	& 1.62$^{+0.18}_{-0.21}$	& 0.14$^{+0.01}_{-0.02}$	& 10.78 & 7.80$-$11.62  \\
\\
	& 4\&5	& 429 & PO	        & 11.84/15	& 0.69	& 2.2$^{+4.0}_{-2.2}$	& 1.85$^{+0.17}_{-0.20}$	& -				& 3.77 & 2.99$-$4.73  \\
	& 4\&5	& 429 & DBB	        & 28.13/16	& 0.03	& 2.8Fr            	& -                         	& 0.88$^{+0.10}_{-0.09}$	& 0.80 & 0.33$-$1.28  \\


\\
S103	& All	& 637	& PO	        & 1272C     & 58\%C	& 11.6$^{+3.3}_{-3.1}$	& 1.79$^{+0.14}_{-0.13}$	& -			 	& 3.89 & 2.88$-$5.65 \\
	& All	& 637	& DBB		& 1282C     & 85\%C	& 2.8Fr				& -				& 1.14$^{+0.08}_{-0.09}$	& 2.78 & 1.92$-$3.65  \\



\enddata
\tablecomments{In column 1 optical correlations (or lack thereof) are indicated by the symbols: $GC$ indicating a GC-LMXB source and $F$ indicating that the LMXB has been confirmed as a field source. No symbol indicates that there are insufficient data to establish the presence of an optical counterpart. When the Cash statistic has been used in preference to \chisq\ a C is included in columns 5 \& 6. A value of 2.8Fr in column 7 denotes that the best-fit model has had the absorption column frozen to the Galactic value. Column 11 presents the range in X-ray luminosity where the lowest value indicates the 1$\sigma$ lower-bound value from the lowest flux individual observation from the joint fit. The higher value indicates the 1$\sigma$ upper-bound value for the most luminous individual observation from the joint fit. For best-fits from individual observations the range indicates the lower- and upper-bound values from that single observation. }
\end{deluxetable}


\end{document}